\documentclass{acm_proc_article-sp}
\usepackage{url}
\usepackage[rgb]{xcolor}
\usepackage[subrefformat=parens,labelformat=parens]{subfig}
\usepackage{caption}
\usepackage{moreverb, amsmath, amsfonts}
\usepackage{graphicx}
\usepackage{placeins}
\usepackage[ruled]{algorithm2e}
\usepackage{booktabs}
\usepackage{multirow}
\usepackage{xspace}
\usepackage{todonotes}
\usepackage{paralist}

\usepackage{adjustbox}
\usepackage{scalefnt}

\usepackage{tikz}

\usepackage{pgf}
\usetikzlibrary{patterns,calc}

\usetikzlibrary{positioning}

\makeatletter

\newenvironment{topbox}[2]{%
    \pgfmathsetlength\@tempdima{#2}%
    \pgfmathsetlength\pgf@yc{\pgfkeysvalueof{/pgf/inner ysep}}%
    \advance\@tempdima by -2\pgf@yc
    \begin{lrbox}{\@tempboxa}%
        \begin{minipage}[t]{#1}%
           \vspace{0pt}%
}{%
        \end{minipage}%
    \end{lrbox}%
    \ifdim\@tempdima>\dp\@tempboxa
        \dp\@tempboxa=\@tempdima
    \fi
    \box\@tempboxa
}

\makeatother

\tikzstyle{block} = [rectangle,fill=none,solid, draw=black, text centered,minimum width=2em, minimum height=2em,inner sep=0pt]
\DeclareRobustCommand\circled[1]{\tikz[baseline=(char.base)]{ \node[shape=circle, fill=white,draw,inner sep=0pt,text width= 1 em ,minimum size = 1 em, text centered] (char) {#1};}}
\tikzstyle{treenode} = [rectangle,fill=white,solid, draw=black, text centered,minimum size=1.5em, text width=1.5em,inner sep=0pt]
\DeclareRobustCommand\supernode[1]{\tikz[baseline=(char.base)]{ \node[shape=rectangle, fill=white,draw,inner sep=0pt,text width= 1 em ,minimum size = 1 em, text centered] (char) {#1};}}

\newcommand{\mapblock}[4]{
    \pgfmathtruncatemacro{\curproc}{ int(mod(#2-1,#3))*#4 + int(mod(#1-1,#4)) +1 };
}

\newcommand{\colorproc}[3]{
  \pgfmathsetmacro{\h}{#1 /(#2*#3)}
  \definecolor{proccolor}{hsb}{\h, 0.5, 0.8}
}

\newcommand{\blk}[6]{
\node[block,minimum width=#2em, minimum height=#3em #6] at (#4em,-#5em) {#1};
}

\newcommand{\dblknew}[4]{
\blk{#1}{#2}{#2}{#3}{#3}{#4}
}

\newcommand{\mdblk}[6]{
    \pgfmathtruncatemacro{\x}{ int(mod(#2-1,#5))*#6 + int(mod(#2-1,#6)) +1 };
 \pgfmathsetmacro{\h}{\x /(#5*#6)}
\definecolor{currentcolor}{hsb}{\h, 0.5, 0.8}
\node[block,minimum width=#3 em, minimum height=#3 em,fill=currentcolor] at (#4 em,-#4 em) {\Proc{\x}};
}

\newcommand{\elublkFill}[5]{
\node[block,draw=none,minimum width=#2em, minimum height=#3em,fill=white,opacity=1] at (#4em,-#5em) {#1};
\draw[draw=black] ($(#4em,-#5em)- 0.5*(#2em,-#3em)$) -- ($(#4em,-#5em)- 0.5*(#2em,#3em)$);
\draw[draw=black] ($(#4em,-#5em) + 0.5*(#2em,#3em)$) -- ($(#4em,-#5em) + 0.5*(#2em,-#3em)$);

\node[block,draw=none,minimum width=#3em, minimum height=#2em,fill=white,opacity=1] at (#5em,-#4em) {#1};
\draw[draw=black] ($(#5em,-#4em)- 0.5*(#3em,-#2em)$) -- ($(#5em,-#4em)- 0.5*(-#3em,-#2em)$);
\draw[draw=black] ($(#5em,-#4em)- 0.5*(#3em,#2em)$) -- ($(#5em,-#4em)- 0.5*(-#3em,#2em)$);

}

\newcommand{\mlublk}[9]{
    \pgfmathtruncatemacro{\x}{ int(mod(#3-1,#8))*#9 + int(mod(#2-1,#9)) +1 };
 \pgfmathsetmacro{\h}{\x /(#8*#9)}
\definecolor{currentcolor}{hsb}{\h, 0.5, 0.8}
 \node[block,minimum width=#4em, minimum height=#5em,fill=currentcolor] at (#6em,-#7em) {\Proc{\x}};

    \pgfmathtruncatemacro{\x}{ int(mod(#2-1,#8))*#9 + int(mod(#3-1,#9)) +1 };
 \pgfmathsetmacro{\h}{\x /(#8*#9)}
\definecolor{currentcolor}{hsb}{\h, 0.5, 0.8}
 \node[block,minimum width=#5em, minimum height=#4em,fill=currentcolor] at (#7em,-#6em) {\Proc{\x}};
}

\newcommand{\ignore}[1]{}

\newcommand{\CS}{\ensuremath{\mathcal C}\xspace}
\newcommand{\JS}{\ensuremath{\mathcal J}\xspace}
\newcommand{\IS}{\ensuremath{\mathcal I}\xspace}
\newcommand{\KS}{\ensuremath{\mathcal K}\xspace}

\newcommand{\Lmat}{\ensuremath{L}\xspace}

\newcommand{\Proc}[1]{\ensuremath{P_{#1}}\xspace}

\newcommand{\etc}{\textit{etc}.\xspace}

\newcommand{\REV}[1]{#1}

\newcommand{\pselinv}{\texttt{PSelInv}\xspace}

\newcommand{\superlu}{\texttt{SuperLU\_DIST}\xspace}

\newcommand{\colbcast}{\texttt{Col-Bcast}\xspace}
\newcommand{\rowreduce}{\texttt{Row-Reduce}\xspace}
\newcommand{\flattree}{\texttt{Flat-Tree}\xspace}
\newcommand{\btree}{\texttt{Binary-Tree}\xspace}
\newcommand{\modbtree}{\texttt{Shifted Binary-Tree}\xspace}

\title{Enhancing the scalability and load balancing of the parallel
selected inversion algorithm via tree-based asynchronous communication}

\numberofauthors{4}

\author{
\alignauthor
Mathias Jacquelin\\
   \affaddr{Lawrence Berkeley National Laboratory}\\
   \email{mjacquelin@lbl.gov}\\
\alignauthor
Lin Lin\\
   \affaddr{University of California Berkeley}\\
   \affaddr{Lawrence Berkeley National Laboratory}\\
   \email{linlin@math.berkeley.edu}\\
\and
\alignauthor
Nathan Wichmann\\
    \affaddr{Cray Inc.}\\
   \email{wichmann@cray.com}\\
\alignauthor
Chao Yang\\
    \affaddr{Lawrence Berkeley National Laboratory}
   \email{cyang@lbl.gov}\\
}

\begin{document}

\maketitle

\nocite{LinLuYingE2009,LinLuYingCarE2009,LinYangLuEtAl2011,LinYangMezaEtAl2011}

\begin{abstract}

We develop a method for improving the parallel scalability of the
recently developed parallel selected inversion algorithm [Jacquelin, Lin
and Yang 2014], named \texttt{PSelInv}, on massively parallel
distributed memory machines.  In the \texttt{PSelInv} method, we compute
selected elements of the inverse of a sparse matrix $A$ that can be
decomposed as $A = LU$, where $L$ is lower triangular and $U$ is upper
triangular. Updating these selected elements of $A^{-1}$ requires
restricted collective communications among a subset of processors within
each column or row communication group created by a block cyclic
distribution of $L$ and $U$.  We describe how this type of restricted
collective communication can be implemented by using asynchronous
point-to-point MPI communication functions combined with a binary tree
based data propagation scheme. Because multiple restricted collective
communications may take place at the same time in the parallel selected
inversion algorithm, we need to use a heuristic to prevent processors
participating in multiple collective communications from receiving too
many messages.  This heuristic allows us to reduce communication load
imbalance and improve the overall scalability of the selected inversion
algorithm.  For instance, when $6,400$ processors are used, we observe
over 5x speedup for test matrices. It also mitigates the performance
variability introduced by an inhomogeneous network topology.


\end{abstract}

\keywords{selected inversion, distributed memory
parallel algorithm, asynchronous data communication, high performance
computation, load balancing}

\section{Introduction}\label{sec:intro}

Collective communication such as broadcast and reduction 
is an ubiquitous type of communication used in many 
parallel programs.  When such communication is required 
among all processors that belong to a communication group
labeled by a communicator, one can use standard message
passing interface (MPI) functions such as 
{\texttt MPI\_Bcast} and {\texttt MPI\_Reduce}.
The MPI libraries available on most of high performance
computers often provide highly efficient implementations of
these functions. These implementations typically make use of
a tree-based algorithm that minimizes the total communication
volume and the number of messages.

However, in some applications, collective communication
is required only among a subset of processors within a 
predefined communication group, and this subset of processors
may change over time. One such application is the pole expansion and selected
inversion
method~\cite{LinLuYingCarE2009,LinYangMezaEtAl2011,LinChenYangEtAl2013} that can be used to accelerate Kohn-Sham 
density functional theory~\cite{KohnSham1965} based electronic structure calculations.
Because the current MPI standard does not support collective
communication among an arbitrary subset of processors, one must
resort to other mechanisms to accomplish such a communication task.

One possible solution is to determine all collective communication
calls that will be needed in advance and the processors involved in 
each one of these calls, set up multiple communication
groups, and use them whenever they are needed. However, 
the total number of communication groups needed (e.g., in 
the selected inversion algorithm) may exceed 
the capacity of the MPI libraries, which is typically around 
several thousands (currently 4,096 on Cray MPI for instance).
Hence the approach of pre-allocating all communicators is not feasible
for all applications.

Another approach is to create communication groups dynamically
as they are needed, and release them when they are no longer
needed. However, this approach typically incurs a significant amount
of overhead that interferes with the asynchronous nature of the 
parallel selected inversion algorithm, and thus limits its parallel
scalability on large scale distributed memory machines.

Yet another solution is to replace the collective 
communication altogether with point-to-point communications.
Although this approach is plausible when each collective 
communication involves only a few processors distributed 
among a small network of processors, it quickly becomes inefficient
when the number of processors involved in the communication 
becomes large.  One main pitfall of this approach is that 
communication load is not well balanced among different processors. 
Such imbalance can severely impair the overall parallel performance. 
Furthermore, when executed on massively parallel
machines that have a hierarchical and inhomogeneous 
network architecture,
such an approach also introduces performance variability.

However, improvement can be made to reduce 
the overall communication cost if we orchestra the point-to-point
communication in such a way that mimics the collective communication
implemented in standard MPI libraries. That is, if we combine a
tree-based algorithm with asynchronous point-to-point sends and receives, 
we can effectively construct dynamic collective communications 
among an arbitrary subset of processors.

In this paper, we demonstrate that the use of the third option can be
quite effective.  However, it needs to be implemented with care
to accommodate a special feature of the selected inversion 
algorithm that allows several restricted collective communications 
to take place at the same time. We present a heuristic that prevent
processors participating in multiple collective communications
from receiving too many messages. We show that this heuristic is
very effective in reducing the amount of load imbalance. 
It also reduces performance variability induced by an inhomogeneous 
network architecture. 

Our paper is organized as follows. In the next section, we 
briefly describe the selected inversion algorithm and its
parallel implementation. We point out the nature of
collective communications required in the parallel implementation
that calls for the implementation of customized broadcast and
reduction operations built on top of asynchronous 
point-to-point communications. We discuss the construction of 
binary trees to propagate data among different processors and
the heuristic for improving communication load balance.
In section~\ref{sec:numerical}, we report the performance improvement
achieved by this technique.
In particular, we show that the implementation of dynamic collective
communication allows the selected inversion method
to scale efficiently beyond 4000 processors.  
The wall clock time can be reduced by more than a factor of three, and the
overall variation in runtime is also reduced.  This improvement
enables us to use selected inversion based electronic structure 
calculation to over 100,000 cores with the pole expansion and selected 
inversion technique~\cite{LinLuYingCarE2009,LinChenYangEtAl2013}.   

\section{Selected Inversion}\label{sec:prelim}

Let $A\in \mathbb{C}^{N\times N}$ be a non-singular sparse matrix.  
We use $A_{i,j}$ to denote the $(i,j)$-th entry of the matrix $A$, 
and $A_{i,*}$ and $A_{*,j}$ to denote the $i$-th row and the $j$-th column
of $A$, respectively.  We are interested in computing 
{\em selected elements} of $A^{-1}$, defined as
\begin{equation}
  \{(A^{-1})_{i,j}\vert \ \ \mbox{for} \ \ 1\le i,j\le N, \ \ 
  \mbox{such that} \ \ A_{i,j}\ne 0 \}.
  \label{eqn:selelem}
\end{equation}
Sometimes, we only need to compute a subset of these selected elements,
for example, the diagonal elements of $A^{-1}$. The most
straightforward way to obtain these selected elements of $A^{-1}$ is to
compute the full inverse of $A$ and then extract the selected elements.
But this is often prohibitively expensive in practice. 
If a sparse $LU$ factorization  of $A$ is available
(or $LDL^{T}$ factorization if $A$ is symmetric) , a more efficient way 
to achieve this goal is to use an algorithm that makes efficient use 
of the sparse $L$ and $U$ factors of $A$.  In such an algorithm, 
which we call selected inversion ({\em SelInv}) some 
additional elements of $A^{-1}$ may need to be computed. However, the overall 
set of nonzero elements that need to be computed often remains a small 
percentage of all elements of $A^{-1}$ due to the sparsity structure of $A$.

The selected inversion algorithm and its variants have been discussed
in a number of publications~\cite{TakahashiFaganChin1973,ErismanTinney1975,CampbellDavis1995,LiAhmedKlimeckDarve2008,LiDarve2012,LiWuDarve2013,HetmaniukZhaoAnantram2013,AmestoyDuffLExcellentEtAl2012,AmestoyDuffLExcellentEtAl2012a,PetersenLiStokbroEtAl2009,LinLuYingE2009,LinLuYingCarE2009,LinYangMezaEtAl2011,LinYangLuEtAl2011}.
We review the basic ingredients of this algorithm in section~\ref{subsec:basic} 
and describe the recently developed parallel algorithm in section~\ref{sec:pselinv}.

\subsection{Sequential algorithm} \label{subsec:basic}

The selected inversion algorithm can be derived as follows. 
Given a 2-by-2 block partitioning of matrix $A$ of the form
\begin{equation}
	A = \begin{pmatrix}
		A_{1,1} & A_{1,2}\\
		A_{2,1} & A_{2,2}
	\end{pmatrix},
	\label{}
\end{equation}
where $A_{1,1}$ is a scalar entry of $A$. 
$A_{1,1}$ can be expressed as a product of two scalars $L_{1,1}$ and $U_{1,1}$.
In particular, we can pick $L_{1,1}=1$ and $U_{1,1}=A_{1,1}$. Then 
\begin{equation}
	A = \begin{pmatrix}
		L_{1,1} & 0\\
		L_{2,1} & I 
	\end{pmatrix}
	\begin{pmatrix}
		U_{1,1} & U_{1,2}\\
		0      & S_{2,2}
	\end{pmatrix}
	\label{eqn:LU2by2}
\end{equation}
where
\begin{equation}
	L_{2,1}=A_{2,1} (U_{1,1})^{-1}, \quad U_{1,2} = (L_{1,1})^{-1}	A_{1,2}.
	\label{}
\end{equation}
The $L$ and $U$ factors are usually directly accessible in a standard
$LU$ factorization, and
\begin{equation}
	S_{2,2} = A_{2,2} - L_{2,1} U_{1,2}	
	\label{}
\end{equation}
is the Schur complement.  Using the decomposition given by 
Eq.~\eqref{eqn:LU2by2}, we can express $A^{-1}$ as
\begin{equation}
\resizebox{.9\linewidth}{!}{$
	A^{-1} = \left(\hspace{-2mm}\begin{array}{cc}
		 \begin{array}{l}
        (U_{1,1})^{-1} (L_{1,1})^{-1} \\
+ (U_{1,1})^{-1} U_{1,2} S^{-1}_{2,2} L_{2,1} (L_{1,1})^{-1} \hspace{-3mm}
 \end{array} & - (U_{1,1})^{-1} U_{1,2} S^{-1}_{2,2}\\
		-S^{-1}_{2,2} L_{2,1} (L_{1,1})^{-1} & S_{2,2}^{-1}
	\end{array}\right)
.
$}
	\label{eqn:Ainv2by2}
\end{equation}
Since $S_{2,2}$ is the same as $S$ here, without
ambiguity $S_{2,2}^{-1}\equiv (S^{-1})_{2,2}$ can be used.
To simplify the notation, we define the normalized $LU$ factors as
\begin{equation}
  \begin{array}{ll}
	\hat{L}_{1,1} = L_{1,1}, & \hat{U}_{1,1}=U_{1,1},\\
	\hat{L}_{2,1} = L_{2,1}(L_{1,1})^{-1}, & \hat{U}_{1,2} = (U_{1,1})^{-1}
	U_{1,2},
  \end{array}
	\label{}
\end{equation}
and Eq.~\eqref{eqn:Ainv2by2} can be equivalently given by 
\begin{equation}
	A^{-1} = \begin{pmatrix}
    (\hat{U}_{1,1})^{-1} (\hat{L}_{1,1})^{-1} + \hat{U}_{1,2}
    S_{2,2}^{-1}
		\hat{L}_{2,1} & - \hat{U}_{1,2} S_{2,2}^{-1} \\
		-S_{2,2}^{-1} \hat{L}_{2,1}  & S_{2,2}^{-1}
	\end{pmatrix}.
	\label{eqn:Ainv2by2normal}
\end{equation}
Let us denote by $\CS$ the set of indices
\begin{equation}
	\{i|\left(L_{2,1}\right)_{i} \ne 0\} \cup 
	\{j|\left(U_{1,2}\right)_{j} \ne 0\},
	\label{}
\end{equation} 
and assume $S_{2,2}^{-1}$ has already been computed. 
From Eq.~\eqref{eqn:Ainv2by2normal} it can be readily observed that
in order to compute the selected elements of 
$\left(A_{2,1}^{-1}\right)_{i} \equiv -\left( S^{-1}_{2,2}
\hat{L}_{2,1} \right)_{i}$ for $i\in \CS$,  
we only need the entries 
\begin{equation}
\left\{\left( S_{2,2}^{-1} \right)_{i,j} | i\in \CS, j\in \CS\right\}.
	\label{eqn:selectentry2x2}
\end{equation}
The same set of entries of $S_{2,2}^{-1}$ are required to compute 
selected entries of $A_{1,2}^{-1} \equiv -\hat{U}_{1,2}
S^{-1}_{2,2}$. 
No additional entries of $S_{2,2}^{-1}$ are needed to complete
the computation of $A_{1,1}^{-1}$, which involves the matrix product 
of selected entries of $\hat{U}_{1,2}$ and $A_{2,1}^{-1}$.
This procedure can be repeated recursively to compute selected elements
of $S_{2,2}^{-1}$ until $S_{2,2}$ is a scalar of size $1$. A pseudo-code for
demonstrating this column-based selected inversion algorithm for symmetric matrix is
given in~\cite{LinYangMezaEtAl2011}.

\begin{algorithm}
  \DontPrintSemicolon
  \caption{Selected inversion algorithm based on $LU$ factorization.}
  \label{alg:selinvlu}

  \KwIn{\begin{tabular}{l} (1) \begin{minipage}[t]{2.5in} The supernode partition of columns of $A$: $\{1,2,...,\mathcal{N}\}$ \end{minipage}\\
        (2) \begin{minipage}[t]{2.5in} A supernodal $LU$ factorization of $A$ with (unnormalized) $LU$ factors $L$ and $U$.  \end{minipage}
        \end{tabular}
        }

   \KwOut{\begin{minipage}[t]{2.5in} Selected elements of $A^{-1}$, i.e. $A^{-1}_{\IS,\JS}$ such
             that $L_{\IS,\JS}$ is not an empty block. \end{minipage}
        } 

	\For{$\KS = \mathcal{N}, \mathcal{N}-1, ..., 1$}{
    \lnl{alg1.step0} Find the collection of indices\;
  	$\CS=\{\IS~|~\IS>\KS,L_{\IS,\KS}\mbox{ is a
	nonzero block}\}\cup \{\JS~|~\JS>\KS, U_{\KS,\JS}\mbox{ is a
	nonzero block}\}$\;
    \lnl{alg1.step1} $\hat{L}_{\CS,\KS}\gets L_{\CS,\KS} (L_{\KS,\KS})^{-1},
    \hat{U}_{\KS,\CS}\gets (U_{\KS,\KS})^{-1} U_{\KS,\CS}$\; 
  }

	\For{$\KS = \mathcal{N}, \mathcal{N}-1, ..., 1$}{
    Find the collection of indices\;
  	$\CS=\{\IS~|~\IS>\KS,L_{\IS,\KS}\mbox{ is a
	nonzero block}\}\cup \{\JS~|~\JS>\KS, U_{\KS,\JS}\mbox{ is a
	nonzero block}\}$\;
	  \lnl{alg1.step2} Calculate $A^{-1}_{\CS,\KS} \gets -A^{-1}_{\CS,\CS}
	  \hat{L}_{\CS,\KS}$\; 
	  \lnl{alg1.step3} Calculate $A^{-1}_{\KS,\KS} \gets U_{\KS,\KS}^{-1}
  	L_{\KS,\KS}^{-1} - \hat{U}_{\KS,\CS} A^{-1}_{\CS,\KS}$\; 
	  \lnl{alg1.step4} Calculate $A^{-1}_{\KS,\CS} \gets - \hat{U}_{\KS,\CS}
	  A^{-1}_{\CS,\CS}$\;
  }
\end{algorithm}

In practice, a column-based sparse factorization and selected
inversion algorithm may not be efficient due to the lack of level 3 BLAS 
operations.  For a sparse matrix $A$, 
the columns of $A$ and the $L$ factor can be
partitioned into supernodes. A supernode is a maximal set of contiguous
columns $\JS=\{j,j+1,\ldots,j+s\}$ of the $L$ factor that have the
same nonzero structure below the $(j+s)$-th row, and the lower
triangular part of $L_{\JS,\JS}$ is dense. This definition can be 
relaxed to limit the maximal number of columns in a supernode (i.e. sets are not necessarily maximal).
With slight abuse of notation, both a supernode index and the set
of column indices associated with a supernode are denoted by uppercase
script letters such as $\IS,\JS,\KS$ \etc.
$A_{\IS,*}$ and $A_{*,\JS}$ are used to
denote the $\IS$-th block row and the $\JS$-th block column of $A$,
respectively. $A_{\IS,\JS}^{-1}$ denotes the $(\IS,\JS)$-th block of the
matrix $A^{-1}$, i.e. $A_{\IS,\JS}^{-1}\equiv (A^{-1})_{\IS,\JS}$.  When
the block $A_{\IS,\JS}$ itself is invertible, its inverse is denoted by
$(A_{\IS,\JS})^{-1}$ to distinguish from $A_{\IS,\JS}^{-1}$.

Using the supernode notation, a pseudo-code for the selected inversion 
algorithm is given in Algorithm~\ref{alg:selinvlu}.  

\subsection{Parallel selected inversion algorithm} \label{sec:pselinv}

\begin{figure}
\centering

\subfloat[A 4-by-3 example of a 2D processor grid]{
\label{fig:pmatrix_grid}
\begin{adjustbox}{width=.25\linewidth}
\begin{tikzpicture}[every node/.style={font=\tiny}]

    \pgfmathtruncatemacro{\Pr}{4};
    \pgfmathtruncatemacro{\Pc}{3};

\mapblock{1}{1}{\Pr}{\Pc}
\colorproc{\curproc}{\Pr}{\Pc}
\blk{\Proc{\curproc}}{2}{2}{1}{1}{,fill=proccolor};
\mapblock{1}{2}{\Pr}{\Pc}
\colorproc{\curproc}{\Pr}{\Pc}
\blk{\Proc{\curproc}}{2}{2}{1}{3}{,fill=proccolor};
\mapblock{1}{3}{\Pr}{\Pc}
\colorproc{\curproc}{\Pr}{\Pc}
\blk{\Proc{\curproc}}{2}{2}{1}{5}{,fill=proccolor};
\mapblock{1}{4}{\Pr}{\Pc}
\colorproc{\curproc}{\Pr}{\Pc}
\blk{\Proc{\curproc}}{2}{2}{1}{7}{,fill=proccolor};

\mapblock{2}{1}{\Pr}{\Pc}
\colorproc{\curproc}{\Pr}{\Pc}
\blk{\Proc{\curproc}}{2}{2}{3}{1}{,fill=proccolor};
\mapblock{2}{2}{\Pr}{\Pc}
\colorproc{\curproc}{\Pr}{\Pc}
\blk{\Proc{\curproc}}{2}{2}{3}{3}{,fill=proccolor};
\mapblock{2}{3}{\Pr}{\Pc}
\colorproc{\curproc}{\Pr}{\Pc}
\blk{\Proc{\curproc}}{2}{2}{3}{5}{,fill=proccolor};
\mapblock{2}{4}{\Pr}{\Pc}
\colorproc{\curproc}{\Pr}{\Pc}
\blk{\Proc{\curproc}}{2}{2}{3}{7}{,fill=proccolor};

\mapblock{3}{1}{\Pr}{\Pc}
\colorproc{\curproc}{\Pr}{\Pc}
\blk{\Proc{\curproc}}{2}{2}{5}{1}{,fill=proccolor};
\mapblock{3}{2}{\Pr}{\Pc}
\colorproc{\curproc}{\Pr}{\Pc}
\blk{\Proc{\curproc}}{2}{2}{5}{3}{,fill=proccolor};
\mapblock{3}{3}{\Pr}{\Pc}
\colorproc{\curproc}{\Pr}{\Pc}
\blk{\Proc{\curproc}}{2}{2}{5}{5}{,fill=proccolor};
\mapblock{3}{4}{\Pr}{\Pc}
\colorproc{\curproc}{\Pr}{\Pc}
\blk{\Proc{\curproc}}{2}{2}{5}{7}{,fill=proccolor};

\end{tikzpicture}
\end{adjustbox}
}
~~~~~~~~
\subfloat[2D block cyclic distribution of \REV{\pselinv sparse matrix}
data structure on a 4-by-3 processor grid]{
\label{fig:pmatrix_layout}
\begin{adjustbox}{width=.45\linewidth}
\begin{tikzpicture}

    \pgfmathtruncatemacro{\Pr}{4};
    \pgfmathtruncatemacro{\Pc}{3};

\node at (-2 em, -1em) {\supernode{1}};
\node at (-2 em, -3em) {\supernode{2}};
\node at (-2 em, -6em) {\supernode{3}};
\node at (-2 em, -9em) {\supernode{4}};
\node at (-2 em, -13em) {\supernode{5}};
\node at (-2 em, -17em) {\supernode{6}};
\node at (-2 em, -19em) {\supernode{7}};
\node at (-2 em, -21.5em) {\supernode{8}};
\node at (-2 em, -24em) {\supernode{9}};
\node at (-2 em, -27em) {\supernode{10}};

\node at (1em ,2 em) {\supernode{1}};
\node at (3em ,2 em) {\supernode{2}};
\node at (6em ,2 em) {\supernode{3}};
\node at (9em ,2 em) {\supernode{4}};
\node at (13em,2 em) {\supernode{5}};
\node at (17em,2 em) {\supernode{6}};
\node at (19em,2 em) {\supernode{7}};
\node at (21.5em,2 em) {\supernode{8}};
\node at (24em,2 em) {\supernode{9}};
\node at (27em,2 em) {\supernode{10}};
\mdblk{}{1}{2}{1}{\Pr}{\Pc};
\mdblk{}{2}{2}{3}{\Pr}{\Pc};
\mdblk{}{3}{4}{6}{\Pr}{\Pc};
\mdblk{}{4}{2}{9}{\Pr}{\Pc};
\mdblk{}{5}{6}{13}{\Pr}{\Pc};
\mdblk{}{6}{2}{17}{\Pr}{\Pc};
\mdblk{}{7}{2}{19}{\Pr}{\Pc};
\mdblk{}{8}{3}{21.5}{\Pr}{\Pc};
\mdblk{}{9}{2}{24}{\Pr}{\Pc};
\mdblk{}{10}{4}{27}{\Pr}{\Pc};

\draw (0em,0em) --   (29em ,0em);
\draw (0em,-2em) --  (29em ,-2em);
\draw (0em,-4em) --  (29em ,-4em);
\draw (0em,-8em) --  (29em ,-8em);
\draw[thick] (0em,-10em) -- (29em ,-10em);
\draw (0em,-16em) -- (29em ,-16em);
\draw (0em,-18em) -- (29em ,-18em);
\draw (0em,-20em) -- (29em ,-20em);
\draw[thick] (0em,-23em) -- (29em ,-23em);
\draw (0em,-25em) -- (29em ,-25em);
\draw (0em,-29em) -- (29em ,-29em);

\draw (0em,0em) --  (0em ,-29em);
\draw (2em ,0em) -- (2em ,-29em);
\draw (4em ,0em) -- (4em ,-29em);
\draw[thick] (8em ,0em) -- (8em ,-29em);
\draw (10em,0em) -- (10em,-29em);
\draw (16em,0em) -- (16em,-29em);
\draw[thick] (18em,0em) -- (18em,-29em);
\draw (20em,0em) -- (20em,-29em);
\draw (23em,0em) -- (23em,-29em);
\draw[thick] (25em,0em) -- (25em,-29em);
\draw (29em,0em) -- (29em,-29em);

\mlublk{}{1}{2}{2}{2}{1}{3}{\Pr}{\Pc};
\mlublk{}{2}{5}{2}{6}{3}{13}{\Pr}{\Pc};

\elublkFill{}{2}{1}{3}{11.5};
\elublkFill{}{2}{1}{3}{15.5};
\draw[draw=black] ($(15.5em,-3em) + 0.5*(1em,2em)$) -- ($(15.5em,-3em) + 0.5*(1em,-2em)$);
\draw[draw=black] ($(3em,-15.5em) + 0.5*(2em,-1em)$) -- ($(3em,-15.5em) + 0.5*(-2em,-1em)$);

\mlublk{}{3}{4}{4}{2}{6}{9}{\Pr}{\Pc};
\mlublk{}{4}{5}{2}{6}{9}{13}{\Pr}{\Pc};

%
\mlublk{}{5}{9}{6}{2}{13}{24}{\Pr}{\Pc};

\mlublk{}{6}{8}{2}{3}{17}{21.5}{\Pr}{\Pc};
\mlublk{}{6}{10}{2}{4}{17}{27}{\Pr}{\Pc};

\elublkFill{}{2}{1}{17}{27.5};
\node at (17em,-27em) {\Proc{6}};
\node at (27em,-17em) {\Proc{4}};

\mlublk{}{7}{8}{2}{3}{19}{21.5}{\Pr}{\Pc};

\mlublk{}{8}{9}{3}{2}{21.5}{24}{\Pr}{\Pc};
\mlublk{+}{8}{10}{3}{4}{21.5}{27}{\Pr}{\Pc};

\elublkFill{}{3}{1}{21.5}{27.5};
\node at (21.5em,-27em) {\Proc{5}};
\node at (27em,-21.5em) {\Proc{10}};
\mlublk{}{9}{10}{2}{4}{24}{27}{\Pr}{\Pc};

%
%
%

\end{tikzpicture}
\end{adjustbox}
}

\caption{Data layout of the \REV{internal sparse matrix}
 data structure used by \pselinv.}
\end{figure}
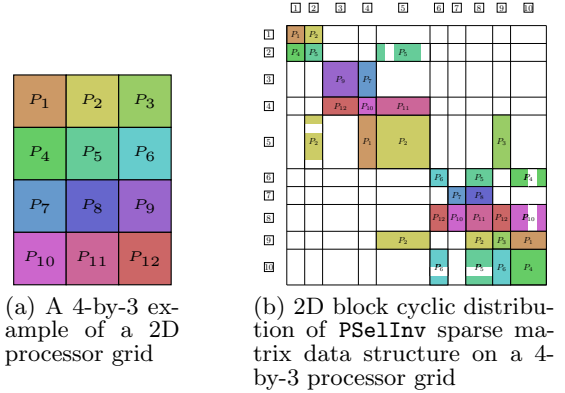

In this section we briefly discuss the parallel implementation of the
selected inversion algorithm, called \texttt{PSelInv}, on distributed
memory parallel machines.  More details of the implementation for
symmetric matrices can be found in~\cite{JacquelinLinYang2014}.  

\pselinv uses the same 2D block cyclic distribution scheme 
employed by \superlu~\cite{LiDemmel2003} to partition and distribute both
the $L$ factor and the selected elements of $A^{-1}$ to 
be computed.
Before the factorization, columns of $A$, $L$ and $U$ are
partitioned into supernodes of various sizes.
This partition is applied to the rows of the input matrix to create
a 2D block partition. These blocks are cyclically mapped onto processors that are 
arranged in a virtual $\mathrm{Pr}$-by-$\mathrm{Pc}$ 2D grid.
The mapping itself does not take the sparsity of the matrix into account,
however only non-zero elements are actually stored.
As an example, a 4-by-3 grid of processors is depicted in
Figure~\subref*{fig:pmatrix_grid}.  
The mapping of the 2D supernode partition of the matrix 
on the 2D processor grid is depicted in Figure~\subref*{fig:pmatrix_layout}.
Each supernodal block column of \Lmat is distributed among processors
that belong to a column of the processor grid.  Each processor may own
multiple matrix blocks.  For instance, nonzero rows in the second 
supernode are owned by processors $\Proc{2}$ and $\Proc{5}$.

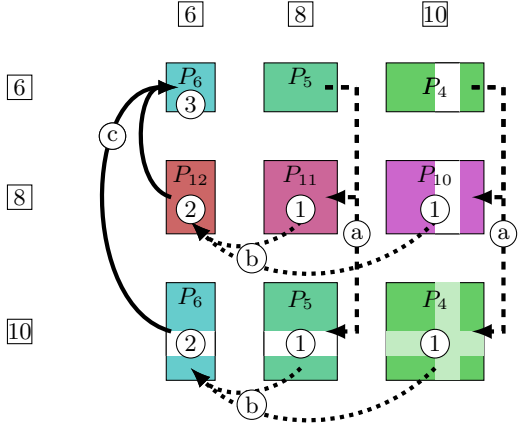
\begin{figure}
\centering
\begin{adjustbox}{scale=1}
\begin{tikzpicture}

    \pgfmathtruncatemacro{\Pr}{4};
    \pgfmathtruncatemacro{\Pc}{3};

\node at (10em, -17em) {\supernode{6}};
\node at (10em, -21.5em) {\supernode{8}};
\node at (10em, -27em) {\supernode{10}};

\node at (17em,-14em) {\supernode{6}};
\node at (21.5em,-14em) {\supernode{8}};
\node at (27em,-14em) {\supernode{10}};

\draw[fill=none,draw=none] (9em,-13em) rectangle +(22em,-18em);

\mapblock{6}{6}{\Pr}{\Pc}
\colorproc{\curproc}{\Pr}{\Pc}
\dblknew{\begin{topbox}{2em}{2em} \vspace{2pt} \center{\Proc{\curproc}} \end{topbox}}{2}{17}{,fill=proccolor};

\mapblock{8}{8}{\Pr}{\Pc}
\colorproc{\curproc}{\Pr}{\Pc}
\dblknew{\begin{topbox}{3em}{3em} \vspace{2pt} \center{\Proc{\curproc}} \end{topbox}}{3}{21.5}{,fill=proccolor};

\mapblock{10}{10}{\Pr}{\Pc}
\colorproc{\curproc}{\Pr}{\Pc}
\dblknew{}{4}{27}{,fill=proccolor};


\mapblock{6}{8}{\Pr}{\Pc}
\colorproc{\curproc}{\Pr}{\Pc}
\blk{\begin{topbox}{2em}{3em} \vspace{2pt} \center{\Proc{\curproc}} \end{topbox}}{2}{3}{17}{21.5}{,fill=proccolor};
\mapblock{8}{6}{\Pr}{\Pc}
\colorproc{\curproc}{\Pr}{\Pc}
\blk{\begin{topbox}{3em}{2em} \vspace{2pt} \center{\Proc{\curproc}} \end{topbox}}{3}{2}{21.5}{17}{,fill=proccolor};

\mapblock{6}{10}{\Pr}{\Pc}
\colorproc{\curproc}{\Pr}{\Pc}
\blk{\begin{topbox}{2em}{4em} \vspace{2pt} \center{\Proc{\curproc}} \end{topbox}}{2}{4}{17}{27}{,fill=proccolor};
\mapblock{10}{6}{\Pr}{\Pc}
\colorproc{\curproc}{\Pr}{\Pc}
\blk{\Proc{\curproc}}{4}{2}{27}{17}{,fill=proccolor};



\mapblock{8}{10}{\Pr}{\Pc}
\colorproc{\curproc}{\Pr}{\Pc}
\blk{}{3}{4}{21.5}{27}{,fill=proccolor};
\mapblock{10}{8}{\Pr}{\Pc}
\colorproc{\curproc}{\Pr}{\Pc}
\blk{}{4}{3}{27}{21.5}{,fill=proccolor};

\mapblock{10}{10}{\Pr}{\Pc}
\colorproc{\curproc}{\Pr}{\Pc}
\blk{}{4}{1}{27}{27.5}{,fill=proccolor!40,draw=none};
\blk{}{1}{4}{27.5}{27}{,fill=proccolor!40,draw=none};

\elublkFill{}{2}{1}{17}{27.5};

\elublkFill{}{3}{1}{21.5}{27.5};
\mapblock{10}{6}{\Pr}{\Pc}
\node at (27em,-17em) {\Proc{\curproc}};

\mapblock{8}{10}{\Pr}{\Pc}
\node at (21.5em,-25.7em) {\Proc{\curproc}};
\mapblock{10}{8}{\Pr}{\Pc}
\node at (27em,-20.6em) {\Proc{\curproc}};

\mapblock{10}{10}{\Pr}{\Pc}
\node at (27em,-25.7em) {\Proc{\curproc}};


\draw[dashed,ultra thick] (22.5em, -17em) -- (23.8em,-17em);
\draw[dashed,ultra thick, -latex] (23.8em, -17em) |- (22.5em,-21.5em);
\draw[dashed,ultra thick] (28.5em, -17em) -- (29.8em,-17em);
\draw[dashed,ultra thick, -latex] (29.8em, -17em) |- (28.5em,-21.5em);

\draw[dashed,ultra thick] (22.5em, -17em) -- (23.8em,-17em);
\draw[dashed,ultra thick, -latex] (23.8em, -17em) |- (22.5em,-27em);
\draw[dashed,ultra thick] (28.5em, -17em) -- (29.8em,-17em);
\draw[dashed,ultra thick, -latex] (29.8em, -17em) |- (28.5em,-27em);

\path (21.5em,-22.5em) edge [dotted,ultra thick, -latex,out=225, in=-45] (17em, -22.5em);
\path (21.5em,-28.5em) edge [dotted,ultra thick, -latex,out=225, in=-45] (17em, -28.5em);
\path (27em,-22.5em) edge [dotted,ultra thick, -latex,out=225, in=-45] (17em, -22.5em);
\path (27em,-28.5em) edge [dotted,ultra thick, -latex,out=225, in=-45] (17em, -28.5em);

\path (16.2em, -21.5em) edge [ultra thick, -latex,out=165, in=-180] (16.5em,-17em);
\path (16.2em, -27em) edge [ultra thick, -latex,out=165, in=-180] (16.5em,-17em);

\node at (29.8 em, -23em) {\circled{a}};
\node at (23.8 em, -23em) {\circled{a}};

\node at (21.5 em, -22em) {\circled{1}};
\node at (27 em, -22em) {\circled{1}};
\node at (21.5 em, -27.5em) {\circled{1}};
\node at (27 em, -27.5em) {\circled{1}};

\node at (19.5 em, -30em) {\circled{b}};
\node at (19.5 em, -24em) {\circled{b}};

\node at (17 em, -22em) {\circled{2}};
\node at (17 em, -27.5em) {\circled{2}};

\node at (13.8 em, -19em) {\circled{c}};
\node at (13.8 em, -19em) {\circled{c}};

\node at (17 em, -17.7em) {\circled{3}};



\end{tikzpicture}
\end{adjustbox}

\caption{Task parallelism and communication pattern for the supernode \supernode{6}.
There are 6 steps: \circled{a} broadcast $\hat{L}$, \circled{1} compute $A^{-1} \hat{L}$, \circled{b} reduce
$A^{-1} \hat{L}$, \circled{2} compute $\hat{L}^T A^{-1} \hat{L}$, \circled{c} reduce
$\hat{L}^T A^{-1} \hat{L}$ and
\circled{3} update $A^{-1}$. }

\label{fig:step3}

\end{figure}

%
%

We execute the first loop of Algorithm~\ref{alg:selinvlu} in a separate pass,
since the data communication required in this step is relatively simple.
The processor that owns the block $L_{\KS,\KS}$ broadcasts it to
all processors that own nonzero blocks
\REV{$L_{\IS,\KS}$} in the supernode $\KS$ within the same processor column.
Each of those processors
performs the triangular solve \REV{$\hat{L}_{\IS,\KS}\equiv L_{\IS,\KS}
(L_{\KS,\KS})^{-1}$} for each nonzero block contained in the set
$\CS$ defined in step~\ref{alg1.step0} of the algorithm.  
Because $L_{\IS,\KS}$ is not used in the subsequent steps of selected
inversion, it is overwritten by $\hat{L}_{\IS,\KS}$.
Since communication is limited to a processor column group
only, multiple supernodes can be processed at the same time.


A more complicated communication pattern, which also turns out to be the
most time consuming step in terms of data communication, is required to complete
step~\ref{alg1.step2} in parallel. 
$A^{-1}_{\CS,\CS}$ and $\hat{L}_{\CS,\KS}$ are generally owned by different processor sets.
In \pselinv, we choose to send the $\hat{L}_{\CS,\KS}$ matrix blocks 
to processors owning the {\em matching} blocks of $A^{-1}_{\CS,\CS}$ to
perform the matrix-matrix multiplication, i.e.
a particular matrix block $\hat{L}_{\IS,\KS}$ 
needs to be communicated to all processors within the same column
group of processors among which $A^{-1}_{\CS,\IS}$ is distributed.

However, since the processor owning $\hat{L}_{\IS,\KS}$ is 
generally not in the same processor row/column group that owns $A^{-1}_{\CS,\IS}$,
the communication cannot be performed by using a simple broadcast
procedure. We briefly describe the communication pattern for this step
here, since most of the communication cost is spent on this step.
In \pselinv, we use point-to-point MPI sends that originate 
from the processor owning $\hat{L}_{\IS,\KS}$ to the group of processors 
holding $A^{-1}_{\CS,\IS}$. 

For symmetric matrices, as soon as $\hat{L}_{\IS,\KS}$ becomes available,
as illustrated above, it is sent to the processor owning
$\hat{U}_{\KS,\IS}$, which is then overwritten by
$\hat{L}_{\IS,\KS}^{T}$. 
Once $\hat{L}_{\IS,\KS}$ has been sent the processor
mapped to the upper triangular part of the matrix,
step~\ref{alg1.step2} of Algorithm~\ref{alg:selinvlu} can be performed.
$\hat{U}_{\KS,\IS}=\hat{L}^{T}_{\IS,\KS}$ is first sent to all
processors within the same processor column that owns
$\hat{U}_{\KS,\IS}$.  The matrix-matrix product
$A^{-1}_{\JS,\IS}\hat{L}_{\IS,\KS}$ is then performed locally on each
processor owning $A^{-1}_{\JS,\IS}$ using BLAS3 kernel.
Then, contributions $A^{-1}_{\JS,\IS}\hat{L}_{\IS,\KS}$ are 
reduced within each processor rows owning $\hat{L}_{\JS,\KS}$. 
The $A^{-1}_{\JS,\KS}$ block in step~\ref{alg1.step2} of
Algorithm~\ref{alg:selinvlu} is now fully computed. 

Figure~\ref{fig:step3} illustrates how this step is completed for 
supernode $\KS=\supernode{6}$. We use circled letters
$\circled{a}, \circled{b}, \circled{c}$ to label 
communication events, and circled numbers
$\circled{1}, \circled{2}, \circled{3}$ to label computational events. 
$\hat{U}_{6,8}=\hat{L}_{8,6}^{T}$ is sent by $P_5$ 
to all processors within the processor column.
This group include both $P_{5}$ and $P_{11}$.  
Similarly $\hat{L}_{10,6}$ is broadcast from $P_4$ to all
other processors within the processor column.
Local GEMMs are then performed on 
$P_{11}$, $P_{10}$, $P_{4}$ and $P_5$ simultaneously, before being
reduced onto $P_{12}$ and $P_{5}$ within their respective processor row.
After this step, $A^{-1}_{8,6}$ and $A^{-1}_{10,6}$ become available on
$P_{12}$ and $P_{6}$ respectively.

In~\cite{JacquelinLinYang2014}, we pointed out that an additional 
coarse-grained level of parallelism exists in the second loop 
of Algorithm~\ref{alg:selinvlu}. Different loop iterates can be
executed simultaneously if 1) there is no data dependency among these 
iterates; 2) there is no overlap among the processors that own data blocks
belonging to these loop iterates.

The absence of data dependency among different loop iterates 
results from the sparsity structure of $A$ and its $L$ and $U$ 
factors, and can be exposed by the elimination tree~\cite{Liu1990} 
associated with a sparse $LU$ factorization. Although two supernodes 
belonging to two different branches of the elimination tree would need 
to communicate with their common ancestors in the selected inversion
algorithm, these communications do not have to take place at the same time.
Thus, it is still possible to process these two supernodes simultaneously.
However, due to the 2D cyclic distribution of the supernodes,
it is possible that some of the matrix blocks belonging to two
independent supernodes are owned by the same processor. In that
case, full parallelism cannot be achieved between the two supernodes.

In \pselinv, we do not explicitly use the MPI\_Barrier function 
for synchronization. The synchronization is only imposed through
data dependencies. As a result, tasks associated with different
supernodes can be executed concurrently if these supernodes are on
different critical paths of the elimination tree, and if there is no
overlap among processors mapped to these critical paths. In this sense,
the asynchronous task formulation tries to achieve two goals:
\textbf{pipelining} computations and \textbf{overlapping} communication
with computations. 

\section{Collective communication in \newline parallel selected inversion}\label{sec:comm}


We can clearly see that the parallelization strategy we presented
in section~\ref{sec:pselinv} involves a fair amount of data communication.
Most of this communication occurs in the second loop of 
Algorithm~\ref{alg:selinvlu}, and in particular, step~\ref{alg1.step4} of 
the algorithm.
Therefore, the performance of our parallel implementation of the 
selected inversion algorithm depends critically on how
$\hat{L}_{\IS,\KS}$ is sent to the matching blocks of $A_{\CS,\CS}^{-1}$ 
and how local products $A^{-1}_{\JS,\IS}\hat{L}_{\IS,\KS}$ are reduced 
to the processor that owns the $\hat{L}_{\JS,\KS}$ block in supernode $\KS$.
These data communications are \textbf{collective} in nature.  
However, they should be carefully treated in the sense that each \textbf{broadcast} 
operation labeled by $\circled{a}$ 
(or \textbf{reduction} operation labeled by $\circled{b}$) in
Figure~\ref{fig:step3} involves only a subset of processors 
within a column or row processor group defined within a 
virtual 2D processor grid shown in Figure~\subref*{fig:pmatrix_layout}.
We will call this type of collective communication {\em restricted
collective communication}. Furthermore, in \pselinv, the subset of 
processors involved in restricted collective communications
varies when different supernodes are processed
due to the general sparsity structure of $L$ and $U$.

As we indicated in the introduction, one way to implement
these collective communication is to construct all possible 
communication groups, each containing a subset of processors
involved in each \textbf{broadcast} and \textbf{reduction}
operation respectively, in advance, and use \texttt{MPI\_Bcast} 
and \texttt{MPI\_Reduce} functions available in standard MPI 
libraries to perform these collective communications. However, 
for large problems, the number of communication groups required 
often exceeds what most MPI libraries can provide. Besides the overhead
for creating a large number of MPI communicators is non-negligible.
Thus, this approach is not viable.

Although it is possible to create these communication groups
dynamically, frequent creation and release of communication
groups tends to result in an excessive amount of overhead. 
Even if this overhead is negligible, using \texttt{MPI\_Bcast}
and \texttt{MPI\_Reduce} is still not optimal because the
{\em collective} and {\em blocking} nature of these functions
reduces the opportunity to exploit the loop level concurrency
available among different supernodes. The subset of ranks involved
in one broadcast may be different from the subset involved in another
broadcast, but it is highly likely that the at least some of the ranks
in one broadcast will also be in another broadcast. Consequently,
the broadcast of one block cannot proceed until the previous
broadcast completes, making the pipelining of updates of different
supernodes in an asynchronous fashion more difficult to achieve. 
Ideally, we would like to have a set of light-weight asynchronous 
\textbf{broadcast} and \textbf{reduction} functions that can
be dynamically created with very little overhead.

We also note that the group of processors involved in each collective
communication is determined once the $L,U$ factors and the 2D processor
mapping is given, and therefore no further communication is needed to
set up the tree once the list of processors is known. With such a list,
the tree structure can be created dynamically with very small overhead.
The buffer arrays for performing the selected inversion is also created
dynamically. 
Such functions are currently not available in standard MPI 
libraries. Therefore, we decided to implement this type of 
restricted collective communication through the use of
point-to-point asynchronous communication functions such
as \texttt{MPI\_Isend} and \texttt{MPI\_Irecv}. 

In the implementation we presented in~\cite{JacquelinLinYang2014},
we simply issue multiple \texttt{MPI\_Isend}'s to send 
$\hat{L}_{\IS,\KS}$ from one processor to other processors 
that own different matching blocks of $A_{\CS,\CS}^{-1}$. 
Similarly, multiple \texttt{MPI\_Irecv}'s are issued to accumulate 
the products $A^{-1}_{\JS,\IS}\hat{L}_{\JS,\KS}$ on the
processor that owns $A_{\JS,\KS}^{-1}$.
By using this simple strategy, we were able to perform 
parallel selected inversion efficiently on $256\sim 1,024$ processors 
depending on the sparsity pattern and the size of a matrix. 
However, when a larger number of processors are used, the performance
of parallel selected inversion quickly deteriorates.  The
performance profile we measured indicated that
communication cost became the dominant cost in those cases.
For instance, for the DG\_PNF14000 matrix in section~\ref{sec:numerical}, 
when a relatively small number of processors $P=256$ is used, the communication
cost is 27\%, and the time spend on the computation, mainly the
matrix-matrix multiplication (GEMM) routine is 73\%. When a large
number of processors $P=4,096$ is used, the communication
cost is 89\%, and the time spend on the computation, mainly the
matrix-matrix multiplication (GEMM) routine is only 11\%.

A closer look at the communication profile reveals that
the increase of communication cost is partly caused by 
a large variation in communication volumes consumed by different
processors.


The communication imbalance is exacerbated by the inhomogeneity of
the communication bandwidth and latency among different 
nodes and processors.  Figure~\subref*{fig:numerDG} shows that as the number of
processors increases, the amount of \textbf{run time variation} also 
increases when we ran the same executable and input matrices multiple
times.  Since \pselinv is a deterministic algorithm, the run time
variation is likely caused by variation in the communication bandwidth
and latency among different combination of nodes and processors and the actual 
mapping between the 2D virtual processor grid and the physical
layout of the processors.

Any network will have different levels of locality, and the realities 
of packaging dictate that the distance between computational nodes will vary and that
subsets of nodes will share routers while other nodes will not.
In most MPI implementations, ranks are assigned so that consecutive ranks
first fill up a node, and then fill the closest node physically, and so on.
It is thus very likely that jobs placed on machines ranks that
are logically close in \texttt{MPI\_COMM\_WORLD} are also physically
close to each other.
Therefore the goal of our broadcast implementation should be to minimize
the amount of data that needs to be transferred at long distance, both 
logically and physically, while at the same time avoiding hot spots in
the network.

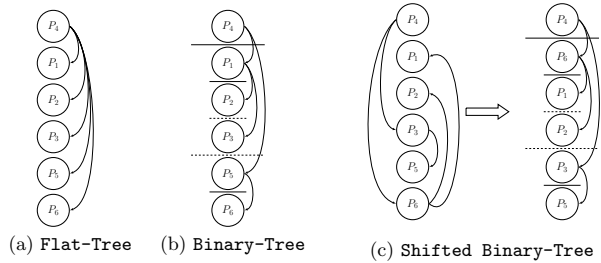
\begin{figure}
\centering
\resizebox{\linewidth}{!}{
\begin{minipage}{1.2\linewidth}
\subfloat[\flattree]{
\label{fig.flat-collectives}
\centering
\parbox{.22\linewidth}{
\centering
\begin{adjustbox}{scale=.3}
\scalefont{2}
\begin{tikzpicture}

\begin{scope}[yscale=-1]

\node[ultra thick,circle,minimum width = 3em,draw=black] (pr) at (0,0) {$P_4$};

\node[ultra thick,circle,minimum width = 3em,draw=black] (pa) at (0,2) {$P_1$};
\node[ultra thick,circle,minimum width = 3em,draw=black] (pb) at (0,4) {$P_2$};
\node[ultra thick,circle,minimum width = 3em,draw=black] (pc) at (0,6) {$P_3$};
\node[ultra thick,circle,minimum width = 3em,draw=black] (pd) at (0,8) {$P_5$};
\node[ultra thick,circle,minimum width = 3em,draw=black] (pe) at (0,10) {$P_6$};

\path (pr.east) edge[thick,out=40,in=-20,looseness =0.9,-latex]  (pa.east);
\path (pr.east) edge[thick,out=40,in=-20,looseness =0.8,-latex]  (pb.east);
\path (pr.east) edge[thick,out=40,in=-20,looseness =0.65,-latex]  (pc.east);
\path (pr.east) edge[thick,out=40,in=-20,looseness =0.55,-latex]  (pd.east);
\path (pr.east) edge[thick,out=40,in=-20,looseness =0.5,-latex]  (pe.east);
\end{scope}

\end{tikzpicture}
\end{adjustbox}
}
}
\subfloat[\btree]{
\label{fig.btree-collectives}
\parbox{.26\linewidth}{
\centering
\begin{adjustbox}{scale=.3}
\scalefont{2}
\begin{tikzpicture}

\begin{scope}[yscale=-1]

\node[ultra thick,circle,minimum width = 3em,draw=black] (pr) at (0,0) {$P_4$};

\node[ultra thick,circle,minimum width = 3em,draw=black] (pn) at (0,2) {$P_1$};
\node[ultra thick,circle,minimum width = 3em,draw=black] (po) at (0,4) {$P_2$};
\node[ultra thick,circle,minimum width = 3em,draw=black] (pq) at (0,6) {$P_3$};
\node[ultra thick,circle,minimum width = 3em,draw=black] (ps) at (0,8) {$P_5$};
\node[ultra thick,circle,minimum width = 3em,draw=black] (pt) at (0,10) {$P_6$};

\draw (-2,1) -- (2,1);
\draw[dashed] (-2,7) -- (2,7);
\draw (-1,3) -- (1,3);
\draw[dashed] (-1,5) -- (1,5);

\draw (-1,9) -- (1,9);

\path (pr.east) edge[thick,out=40,in=-20,looseness =0.9,-latex]  (pn.east);
\path (pn.east) edge[thick,out=40,in=-20,looseness =0.8,-latex]  (po.east);
\path (pn.east) edge[thick,out=40,in=-20,looseness =0.65,-latex]  (pq.east);
\path (pr.east) edge[thick,out=40,in=-20,looseness =0.55,-latex]  (ps.east);
\path (ps.east) edge[thick,out=20,in=-20,looseness =0.8,-latex]  (pt.east);
\end{scope}

\end{tikzpicture}
\end{adjustbox}
}
}
\subfloat[\modbtree]{
\label{fig.modbtree-collectives}
\parbox{.5\linewidth}{
\centering
\begin{adjustbox}{scale=.3}
\scalefont{2}
\begin{tikzpicture}

\begin{scope}[yscale=-1, xshift=-14em]

\node[ultra thick,circle,minimum width = 3em,draw=black] (pr) at (0,0) {$P_4$};

\node[ultra thick,circle,minimum width = 3em,draw=black] (pn) at (0,10) {$P_6$};
\node[ultra thick,circle,minimum width = 3em,draw=black] (po) at (0,2) {$P_1$};
\node[ultra thick,circle,minimum width = 3em,draw=black] (pq) at (0,4) {$P_2$};
\node[ultra thick,circle,minimum width = 3em,draw=black] (ps) at (0,6) {$P_3$};
\node[ultra thick,circle,minimum width = 3em,draw=black] (pt) at (0,8) {$P_5$};

\path (pr.west) edge[thick,out=-220,in=-160,looseness =0.6,-latex]  (pn.west);
\path (pn.east) edge[thick,out=40,in=-20,looseness =0.8,-latex]  (po.east);
\path (pn.east) edge[thick,out=-40,in=20,looseness =0.65,-latex]  (pq.east);
\path (pr.west) edge[thick,out=-220,in=-160,looseness =0.55,-latex]  (ps.west);
\path (ps.east) edge[thick,out=20,in=-20,looseness =0.8,-latex]  (pt.east);
\end{scope}

\begin{scope}[yscale=-1]

\draw[ultra thick] (-9em,4.8) -- (-6em,4.8) -- (-6em,4.6) -- (-5em,5) -- (-6em,5.4) -- (-6em,5.2) -- (-9em,5.2) -- (-9em,4.8);

\end{scope}

\begin{scope}[yscale=-1]

\node[ultra thick,circle,minimum width = 3em,draw=black] (pr) at (0,0) {$P_4$};

\node[ultra thick,circle,minimum width = 3em,draw=black] (pn) at (0,2) {$P_6$};
\node[ultra thick,circle,minimum width = 3em,draw=black] (po) at (0,4) {$P_1$};
\node[ultra thick,circle,minimum width = 3em,draw=black] (pq) at (0,6) {$P_2$};
\node[ultra thick,circle,minimum width = 3em,draw=black] (ps) at (0,8) {$P_3$};
\node[ultra thick,circle,minimum width = 3em,draw=black] (pt) at (0,10) {$P_5$};

\draw (-2,1) -- (2,1);
\draw[dashed] (-2,7) -- (2,7);
\draw (-1,3) -- (1,3);
\draw[dashed] (-1,5) -- (1,5);

\draw (-1,9) -- (1,9);

\path (pr.east) edge[thick,out=40,in=-20,looseness =0.9,-latex]  (pn.east);
\path (pn.east) edge[thick,out=40,in=-20,looseness =0.8,-latex]  (po.east);
\path (pn.east) edge[thick,out=40,in=-20,looseness =0.65,-latex]  (pq.east);
\path (pr.east) edge[thick,out=40,in=-20,looseness =0.55,-latex]  (ps.east);
\path (ps.east) edge[thick,out=20,in=-20,looseness =0.8,-latex]  (pt.east);
\end{scope}

\end{tikzpicture}
\end{adjustbox}
}
}
\end{minipage}
}

\caption{Various possible tree-based communication patterns for the
broadcast operation.}
\label{fig.collectives}
\end{figure}

To reduce communication imbalance and consequently the communication 
cost, we modify the way the collective communication in 
step 3 of Algorithm~\ref{alg:selinvlu} is implemented. Instead of using
a ``centralized'' sender/receiver model for the broadcast and
reduction in which the communication path can be described a \flattree
shown in Figure~\subref*{fig.flat-collectives}, 
we use a binary-tree based algorithm commonly employed in the 
implementation of \texttt{MPI\_Bcast} and \texttt{MPI\_Reduce}.
Compared to the \flattree model, which puts a heavy load on the 
root, the \btree based scheme reduces the total
volume {sent/received} from the root from $p-1$ messages to two messages, 
and spreads the total communication volume among more processors.  
Figure~\subref*{fig.flat-collectives}
and~\subref*{fig.btree-collectives}shows 
how messages are passed among different processors in a \flattree and 
\btree based broadcast operation.

In order to implement a non-blocking \btree based collective communication 
scheme, destination processor of each message has to be specified
in a hierarchical fashion.  We will refer to processors that 
lie between the root and the leaves of the tree as ``internal nodes''. 
Such processors serve as the forwarding processors.
The \btree is built by repeatedly splitting the ordered list of ranks
in two partitions, and chose the first rank in each half to be the
internal nodes at the current level.

The \btree has two main benefits.  First, the large reduction in 
the number of messages sent from the root greatly reduces the chances of an 
instantaneous hot spot in the network around the root node.  Second, 
the \btree greatly increases the chance that data is exchanged between
two ranks that are logically closer, and thus likely physically closer in the network,
by putting them in the same partition.
As an example, Figure~\subref*{fig.btree-collectives} shows that the processors
$P_{1}-P_{6}$ are involved in a broadcast operation, with $P_{4}$ being
the root.  The \flattree simply sends data from $P_{4}$ to all
processors other than $P_{4}$. The \btree uses a pre-designed ordering,
i.e. $P_{4}$ first sends to $P_{1}$ and $P_{5}$ which are of
distance $1$ from the root $P_{4}$.  The data is further broadcast from
$P_{1}$ to $P_{2},P_{3}$, and from $P_{5}$ to $P_{6}$.  The data
communication can be performed recursively for deeper trees.

The drawback of the \btree is that due to the pipelining of different loop iterates in the 
outer-loop of Algorithm~\ref{alg:selinvlu}, each processor may participate 
in several non-blocking restricted collective communication simultaneously. If
that processor is an internal node in several binary communication trees,
the total volume from those many broadcasts passing through that processor can be 
much larger than that sent by other processors.
One can see that with this scheme, the highest numbered rank in a 
column will never be chosen as an internal node and thus will 
never forward any data.
 On the other hand, the lowest numbered
rank in a column will always be chosen as an internal node and thus
will always forward data.  While the exact ranks chosen as internal
nodes will vary depending on the root, patterns of communication 
intensity will develop throughout the range of ranks.  Such a 
striped pattern is clearly seen in the communication volume heat map seen in 
Figure~\subref*{fig:hm_sender_bcastU_btree}.

To alleviate this problem, we use a heuristic method that involves
applying a random circular shift to the list of receiving processor ranks.
Such a procedure, referred to as \modbtree, is depicted in 
Figure~\subref*{fig.modbtree-collectives}. A random position is selected in
the sorted list of ranks and a circular shift is then executed around this position.
The random shift makes it therefore less likely that same processors
will be picked as internal nodes when building multiple binary trees. The
rationale of this circular shift is therefore to smooth the total communication
load across all processors.

In our example in Figure~\subref*{fig.modbtree-collectives}, the
\modbtree \linebreak breaks the pre-designed and monotonically increasing ordering 
of ranks involved in the tree,
and picks a random processor
other than the root ($P_{4}$) to be the first child.  The rest of the
processors follow circularly, so that the sequence
$P_{4}$,$P_{6}$,$P_{1}$,$P_{2}$,$P_{3}$,$P_{5}$ can be regarded as a re-ordered
list to generate the \btree. 
This results in a different data communication pattern.

One can also consider using a fully random permutation of processor ranks. However,
such a permutation would reduce network locality by putting ranks which are logically
``closer'' far from each other. Moreover, our experiments show that this
approach leads to deteriorated load balancing in terms of communication
volume compared to \modbtree.




\begin{figure*}[bthp]
\centering
\subfloat[\flattree]{
\label{fig:hist_sender_bcastU_ftree}
\begin{adjustbox}{scale=0.5}
  \includegraphics{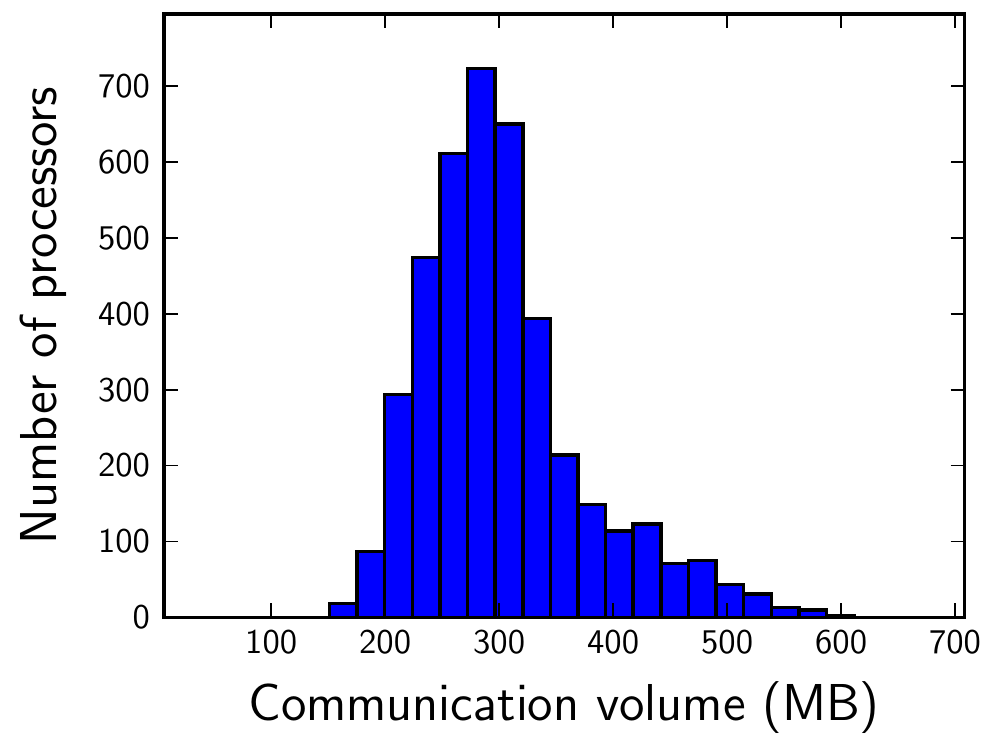}
\end{adjustbox}
}~~~
\subfloat[\btree]{
\label{fig:hist_sender_bcastU_btree}
\begin{adjustbox}{scale=0.5}
  \includegraphics{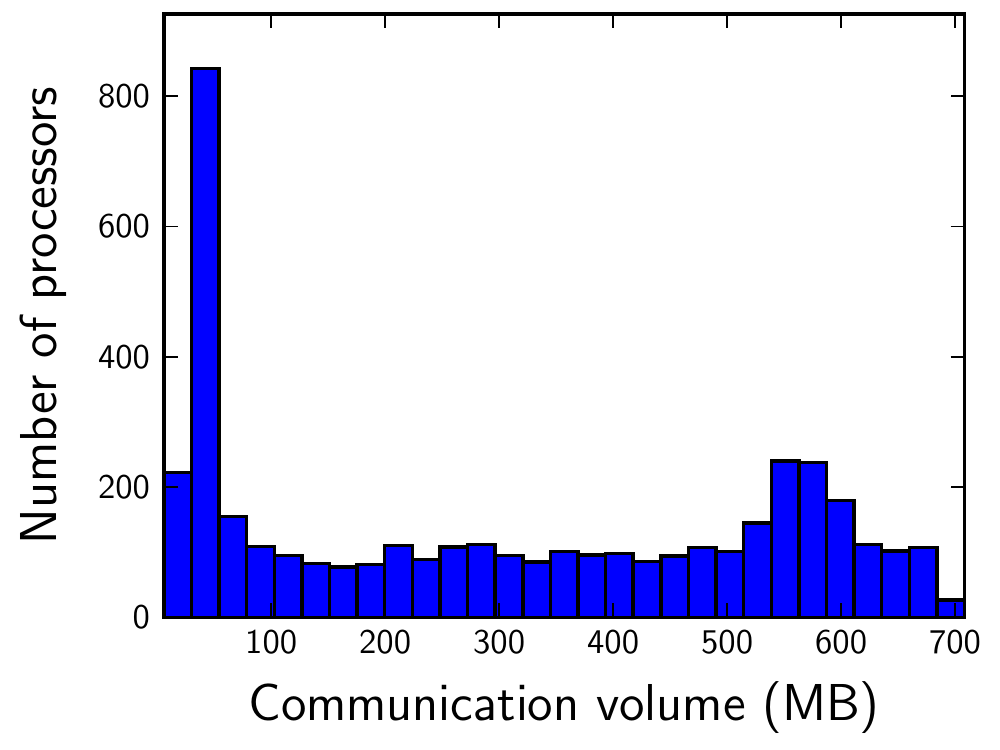}
\end{adjustbox}
}~~~
\subfloat[\modbtree]{
\label{fig:hist_sender_bcastU_modbtree}
\begin{adjustbox}{scale=0.5}
  \includegraphics{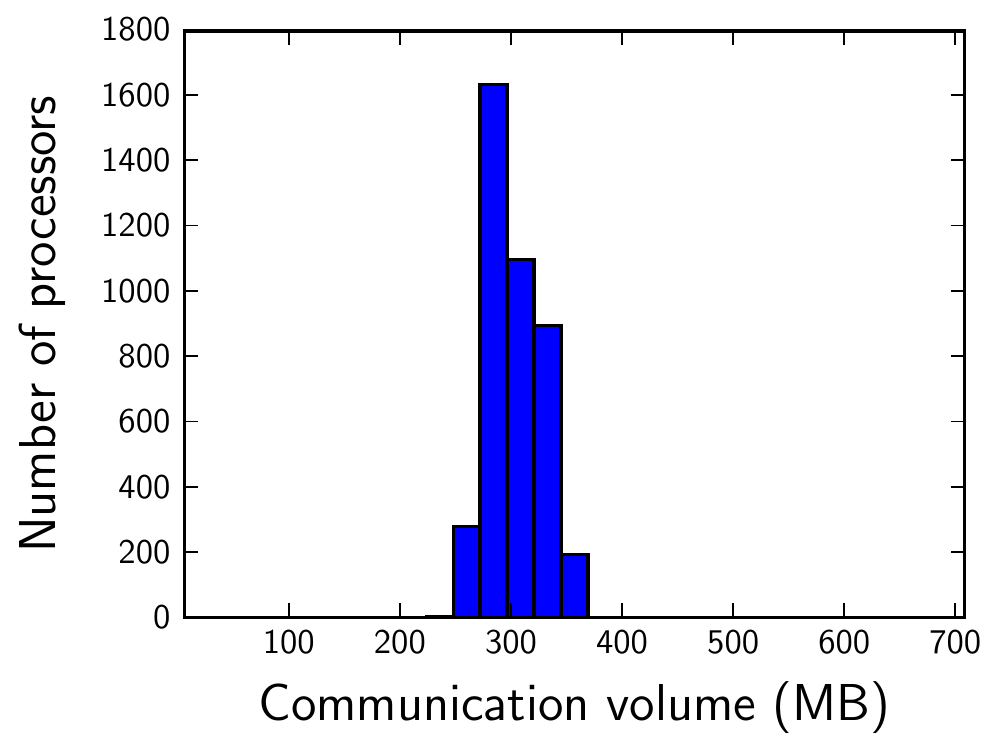}
\end{adjustbox}
}

\caption{Communication volume distribution of \colbcast}
\label{fig:hist_sender_bcastU}
\end{figure*}

Both these binary tree structures do an excellent job of reducing ``hot spots'' in 
the network and reducing the communication distance for data transfer.
As already discussed, the simple \btree structure reduces the number of
messages transferred to/from the root from $p-1$ to just two, greatly 
reducing the chance of an instantaneous hot spot developing at the 
time of that broadcast.  It also increases the chances that data will 
be transferred between ranks that are logically close to each other, 
rather than all data being transferred from the root to the leaf no 
matter the distance between the leaf and the root.  

The \modbtree further reduces the chances of hot spots in
the network given that there are many broadcasts happening
simultaneously.  The communication heat map in
Figure~\subref*{fig:hm_sender_bcastU_modbtree} clearly shows that the
communication volume spreads much more evenly across all ranks. 
The maximum amount of data sent by any rank is also reduced. Note that the circular
shift potentially reduces network locality by putting the highest
rank in the list before the lowest rank. However, this
does not negatively impact the algorithm in any
significant way, since the root and the next level of internal nodes
were not guaranteed to be close to each other when the number of
processors involved in a communication is relatively large.

\section{Numerical results}\label{sec:numerical}
We now report the outcome of a number of computational experiments 
conducted to analyze the communication volume and pattern of
\pselinv, and to evaluate and compare the efficiency of different
ways to implement the restricted collective communication pattern
required in \pselinv.

In all of our experiments, we used the NERSC Edison platform with Cray
XC30 nodes. Each node has 24 cores partitioned among two
Intel Ivy Bridge processors.  Each 12-core processor runs at 2.4GHz. A
single node has 64GB of memory, providing more than 2.6 GB of memory per
core.  

To evaluate the performance of \pselinv, we use two matrices of different 
sizes and sparsity patterns. The \linebreak DG\_PNF14000 matrix is generated
from the electronic structure calculation of a 2D phosphorene nanoflake with
$14,000$ atoms. The matrix is a discretized Kohn-Sham Hamiltonian
obtained from an adaptive local basis expansion scheme combined with a
discontinuous Galerkin framework~\cite{LinLuYingE2012}. This
matrix is relatively dense. The matrix size is 512,000,  with 0.2\% nonzeros
in $A$ and 1.3\% nonzeros in the $L$ and $U$ factors.  
The second matrix is named audikw\_1 matrix obtained from the University
of Florida matrix collection~\cite{FloridaMatrix}. This
matrix is relatively sparse. The matrix size is 943,695,
with 0.009\% nonzeros in $A$ and 0.3\% nonzeros in the $L$ and $U$ factors.  
These two matrices represent two different scenarios in terms of total communication
volume.  For the DG\_PNF14000 matrix, the communication volume of is 
expected to be very large. This can lead to imbalanced data communication 
on different processors.  For the audikw\_1 matrix, the scalability of 
\pselinv is more limited by a larger communication over computation 
ratio.  


\subsection{Communication load analysis}

The first set of experiments aims at analyzing the communication load
among different processors, and comparing the efficiency of using 
different types of
tree structures to implement restricted collective communications
by using asynchronous MPI functions. We report the total data volume sent
from each processor on a $64$-by-$64 = 4,096$ processor grid 
for the audikw\_1 matrix.

Our main focus is the broadcast of blocks of $\hat{L}^{T}_{\IS,\KS}$
within each column group, and the reduction of
$A^{-1}_{\JS,\IS}\hat{L}_{\IS,\KS}$ within each row group. 
These two operations are the most expensive communication steps
of \pselinv, and we refer to them as \colbcast and \rowreduce respectively.
We report the minimum and maximum outgoing volume of data among 
all processors in Table~\ref{tab:colbcast_stats} for different types of
tree-based collective communication schemes.



\begin{figure*}[btph]
\centering
\subfloat[\flattree]{
\label{fig:hm_sender_bcastU_ftree}
\begin{adjustbox}{scale=0.32}
  \includegraphics{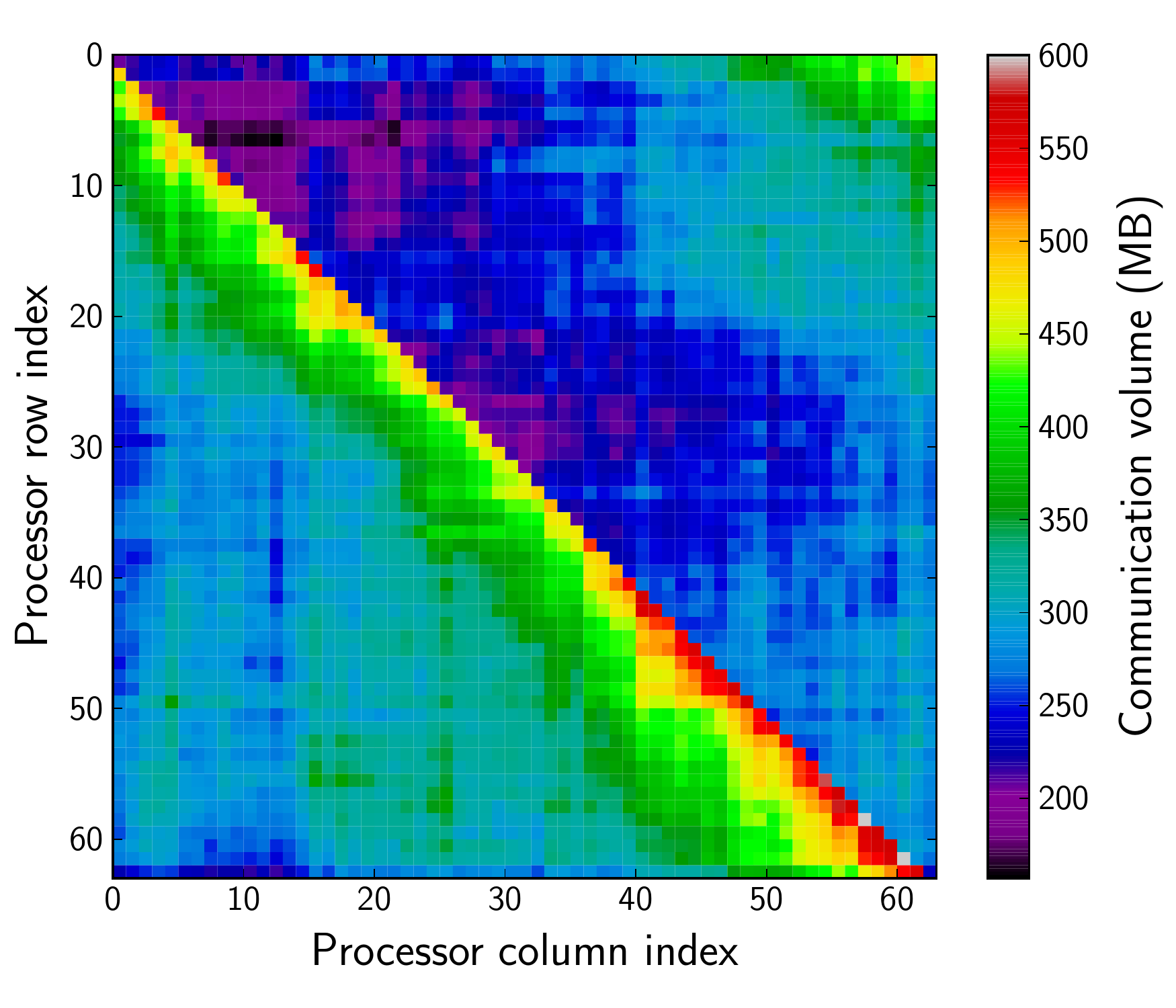}
\end{adjustbox}
}
\subfloat[\btree]{
\label{fig:hm_sender_bcastU_btree}
\begin{adjustbox}{scale=0.32}
  \includegraphics{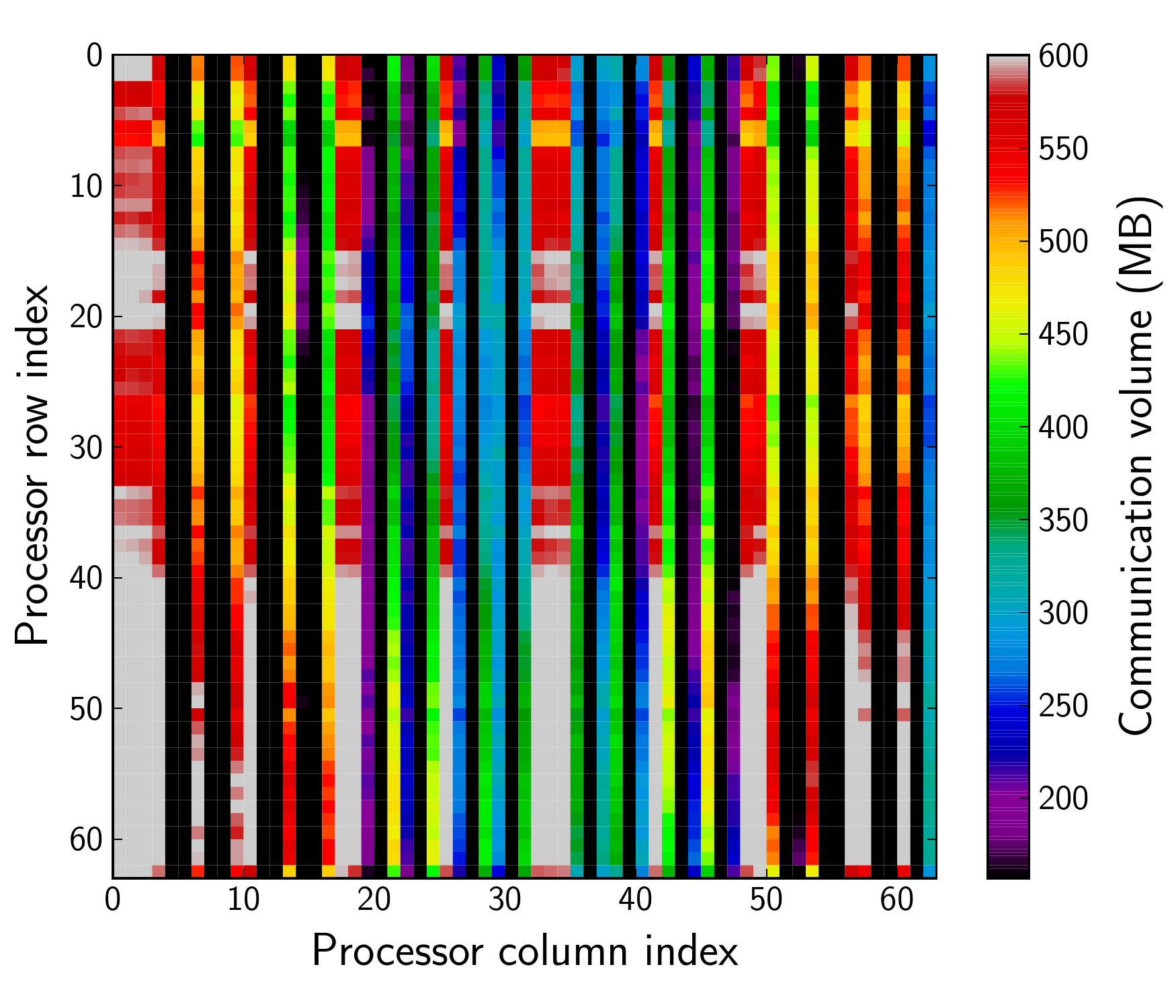}
\end{adjustbox}
}
\subfloat[\modbtree]{
\label{fig:hm_sender_bcastU_modbtree}
\begin{adjustbox}{scale=0.32}
  \includegraphics{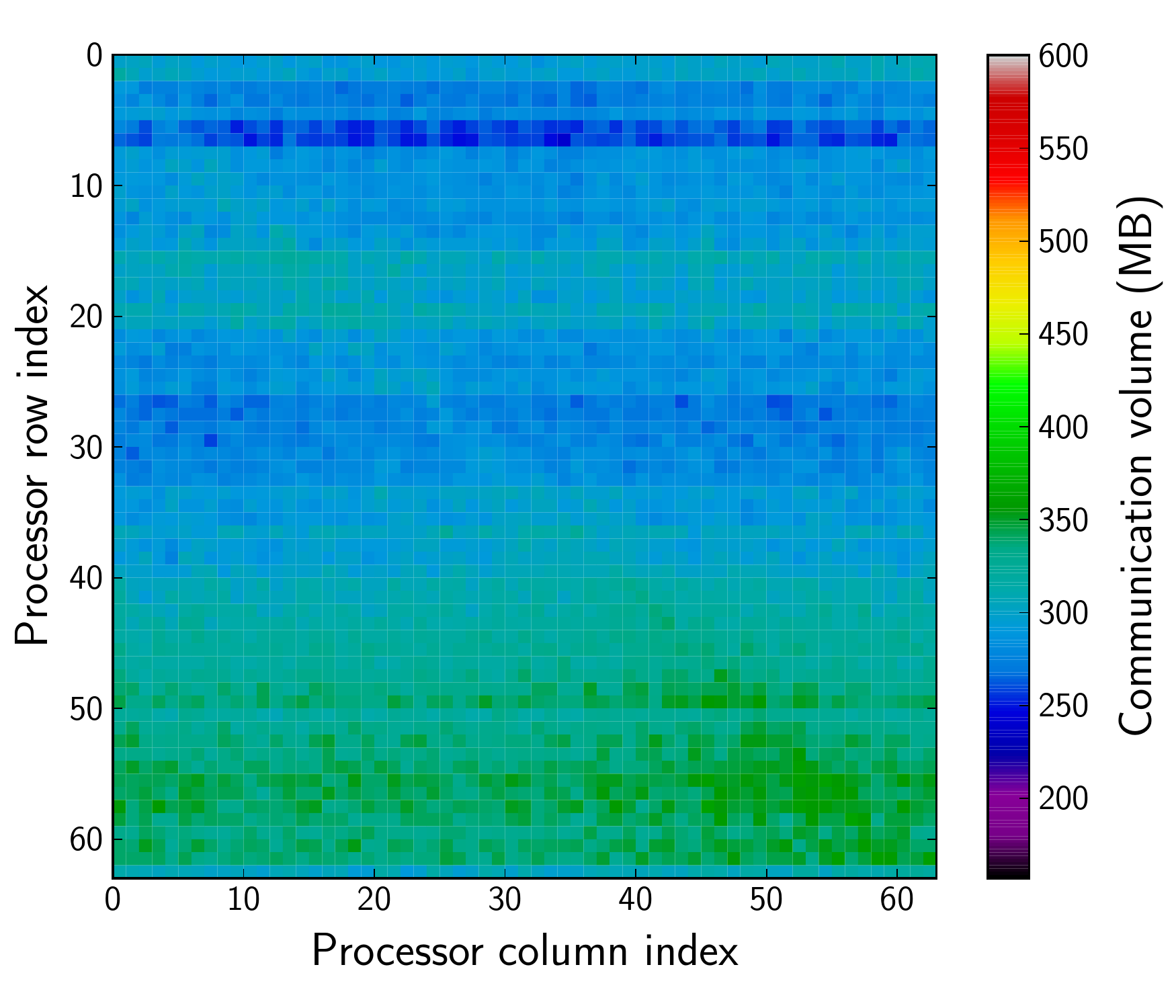}
\end{adjustbox}
}
\caption{Communication volume heat map of \colbcast.}
\end{figure*}

In Figure~\subref*{fig:hm_sender_bcastU_ftree}, we report the volume sent during
\colbcast using the \flattree pattern, as used in the \pselinv
developed in~\cite{JacquelinLinYang2014} (currently released under the
PEXSI package v0.7.3, referred to as \pselinv v0.7.3 below). 
We observe that the data volume associated with processors 
near the diagonal of the 2D processor grid is significantly higher than 
those associated with off-diagonal processors. Significant variation 
of communication volume can also be seen among the diagonal processors
themselves.  Furthermore, from the distribution of the processor load shown
in Figure~\subref*{fig:hist_sender_bcastU_ftree}, we observe that 
some processors send more than twice the average volume of data 
sent by all processors.  This load imbalance creates contention on the
network and limits the strong scalability of \pselinv.




\begin{table}[h]
\centering
\begin{adjustbox}{width=.90\linewidth}
\begin{tabular}{|c|c|c|c|c|}
\hline
Communication tree&  Min& Max & Median& Std. dev\\
\hline
\flattree & 156.951& 600.168&  291.595& 72.048\\
\btree & 5.89874& 708.268&  288.851& 226.565\\
\modbtree & 238.647& 363.336& 298.58& 24.0957\\
\hline
\end{tabular}
\end{adjustbox}
\caption{Volume sent during \colbcast (in MB) for the audikw\_1 matrix. }\label{tab:colbcast_stats}
\end{table}

When a simple \btree is used to organize the way messages are sent
from the root to other processors involved in the restricted collective
communication, load imbalance can still be observed from the heat map 
given in Figure~\subref*{fig:hm_sender_bcastU_btree}. It can be seen from 
Table~\ref{tab:colbcast_stats} and Figure~\subref*{fig:hist_sender_bcastU_btree} that the maximum communication volume among all processors and 
the standard deviation are actually higher than those in the \flattree based
communication. Although
most of the nodes see their load decreased and the median communication
volume reduced from 291 MB to 288 MB, ranks in the last quartile of the most loaded processors
send more than 536 MB of data instead of the 331 MB of data sent using a
\flattree based scheme.  This observation confirms that some processors
participate in multiple \btree based broadcasts as internal nodes.


In order to reduce the likelihood of a processor being chosen repeatedly 
as an internal node, we introduced the \modbtree communication
scheme in Section~\ref{sec:comm}. The communication volume heat map 
associated with this scheme is shown in 
Figure~\subref*{fig:hm_sender_bcastU_modbtree}. 
We use the same colorbar that we used for 
Figure~\subref*{fig:hm_sender_bcastU_ftree} so that we can compare these 
two heat maps directly.  We can clearly see that the overall heat map 
is much ``cooler''.  The communication ``hot spots'' appear to be
eliminated by the \modbtree.
As we explained in section~\ref{sec:comm}, the reduction in 
the overall communication load and the removal of the ``hot spots" 
result from shifting processor ranks in such a way that
different processors are picked as internal nodes of different
communication trees. The effect of this is clearly observed in practice
on the minimum and maximum volumes, given in
Table~\ref{tab:colbcast_stats}.  The variation of the communication
volume among different processors is significantly reduced (resp. MIN
238 MB and MAX 363 MB) than that using \flattree (resp. MIN 156 MB and
MAX 600 MB). The standard deviation is significantly reduced from 72 MB
to 24 MB, confirming the efficiency of the approach.

\begin{figure}[htbp]
\centering
\begin{adjustbox}{width=.50\linewidth}
  \includegraphics{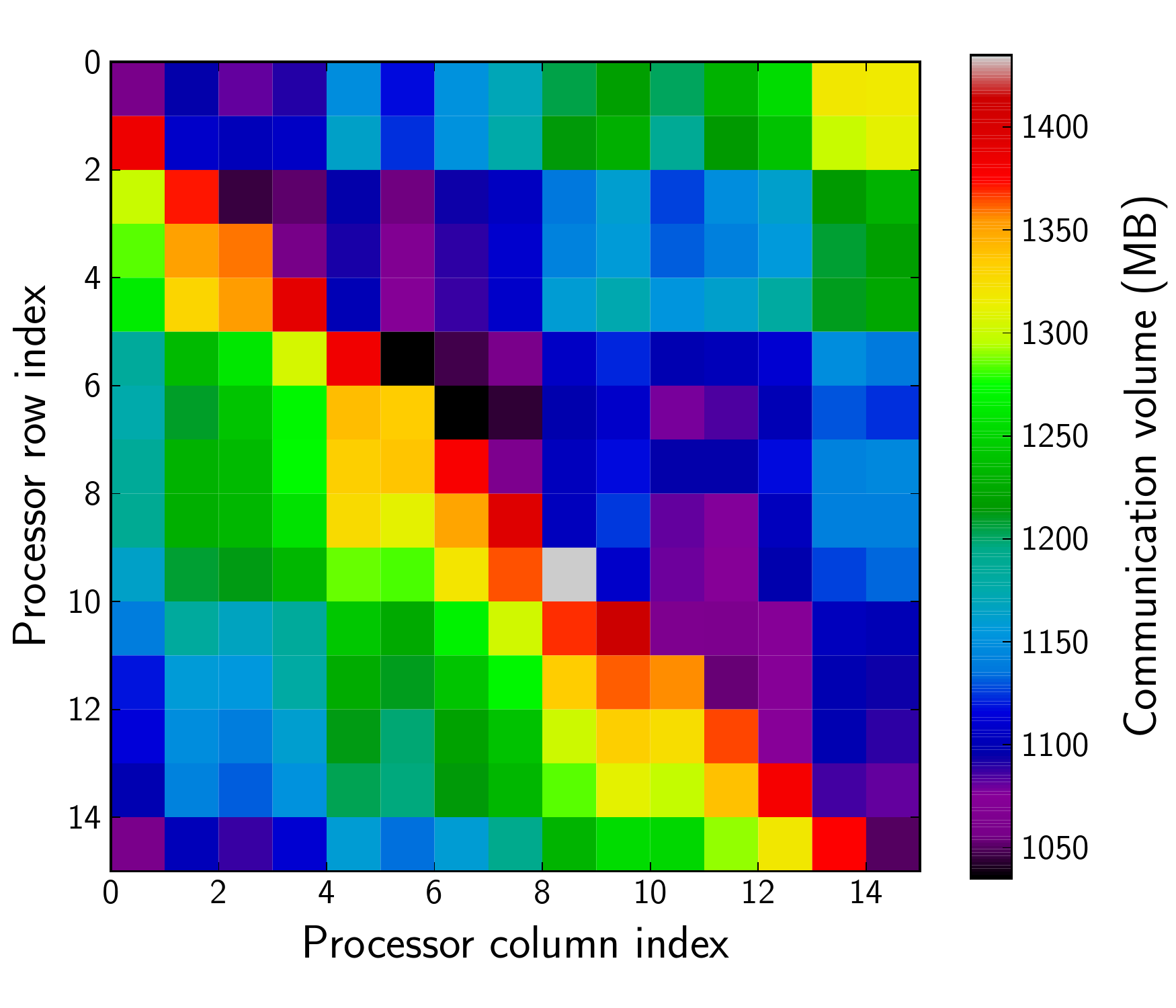}
\end{adjustbox}
\caption{Communication volume heat map of \colbcast using \flattree on 256 processors}
\label{fig:hm_sender_bcastU_ftree_16x16}
\end{figure}


When fewer processors are used to perform \pselinv, the communication 
load imbalance might not be so severe.  
Figure~\ref{fig:hm_sender_bcastU_ftree_16x16} depicts the communication
volume heat map of \colbcast for the same audikw\_1 matrix running
on a 16-by-16 processor grid using the \flattree scheme.
In this case, the average volume is 1185.77 MB, while the standard deviation 
is 96.02 MB, corresponding to 8\% of the average. 
This is sharply lower than the 23.7\% standard deviation 
when \pselinv is carried out on 4,096 processors. 

\begin{figure}[htbp]
\centering
\subfloat[\flattree]{
\label{fig:hm_receiver_reduceL_ftree}
\begin{adjustbox}{width=.49\linewidth}
  \includegraphics{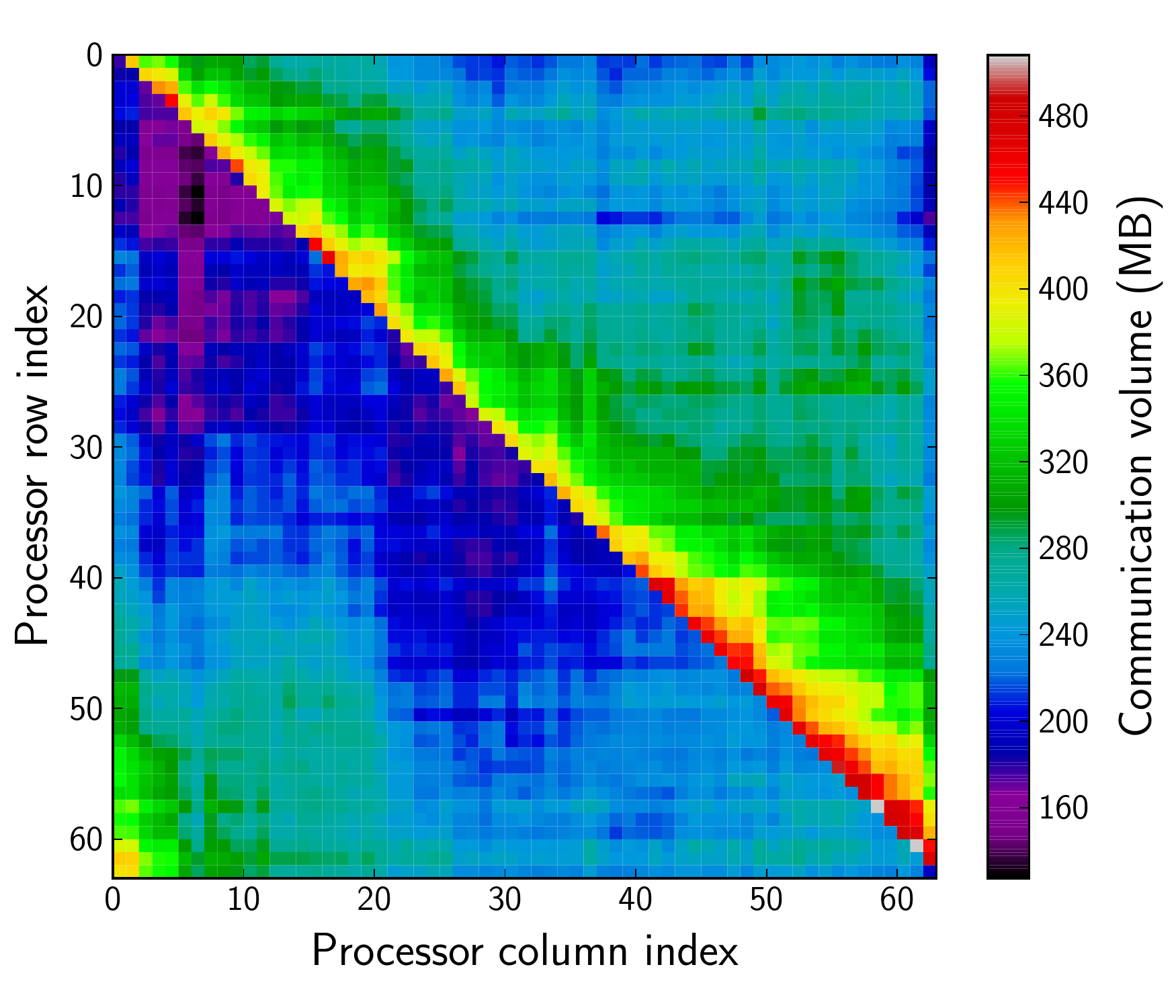}
\end{adjustbox}
}
\subfloat[\modbtree]{
\label{fig:hm_receiver_reduceL_modbtree}
\begin{adjustbox}{width=.49\linewidth}
  \includegraphics{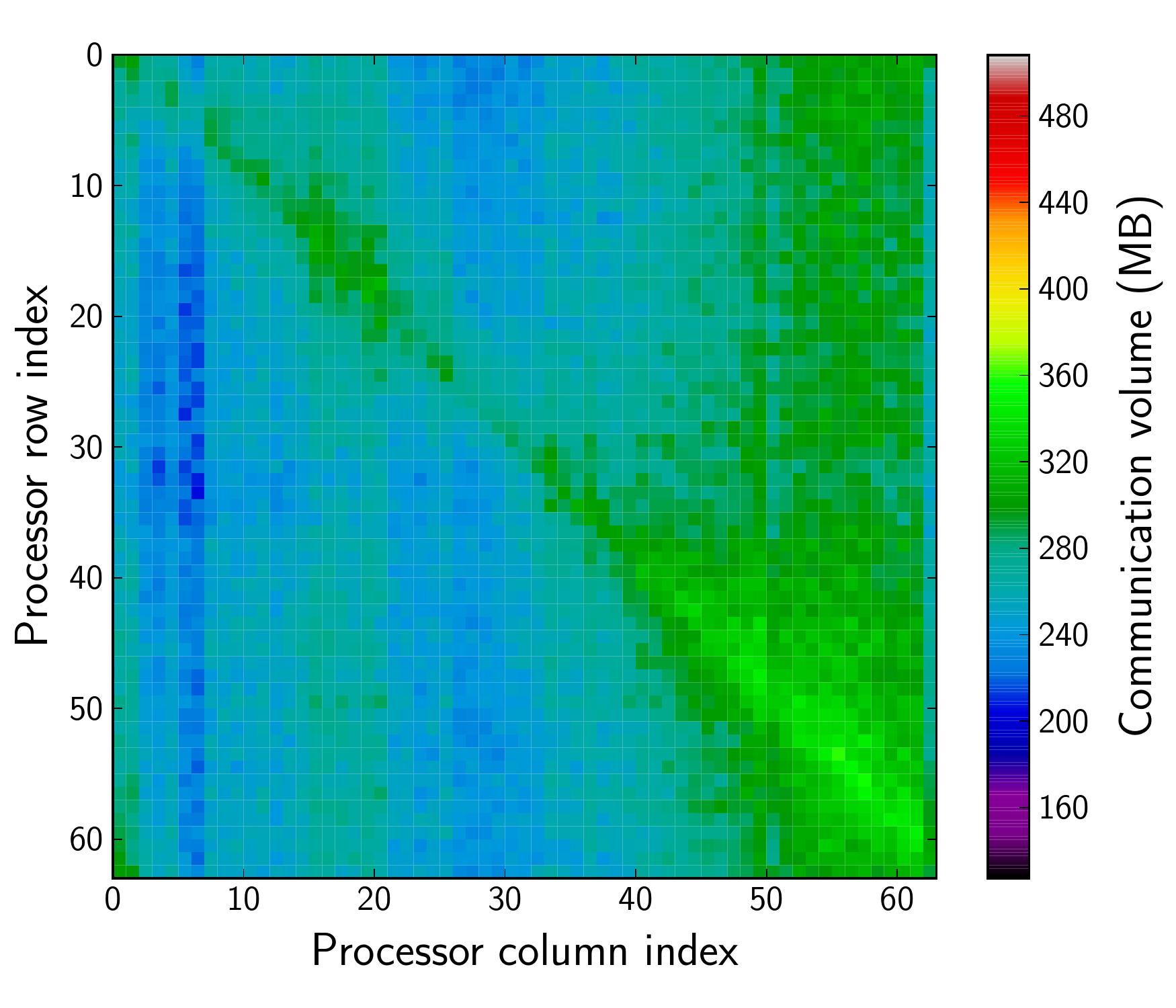}
\end{adjustbox}
}
\caption{Communication volume heat map of \rowreduce }
\label{fig:hm_receiver_reduceL}
\end{figure}

The \rowreduce operation can be seen as the reverse operation of a
broadcast. In this case, it is the amount of data \textit{received}
by each processor that we are concerned with. 
Heat maps corresponding to \flattree and \modbtree
are shown in Figures~\subref*{fig:hm_receiver_reduceL_ftree}
and \subref*{fig:hm_receiver_reduceL_modbtree} respectively. 
We can clearly see that the \modbtree scheme results in 
a more balanced communication load distribution among 
all processors.



%
%

Altogether, our experimental results demonstrate that the use of a binary tree
to organize messages in a restricted collective communication does mitigates 
the inherent load imbalance of the \flattree communication pattern.  
However, since multiple restricted collective communications may take place
at the same time with some of the processors participating in all of them,
the binary trees associated with these collective calls have to be built 
in such a way that these processors are not always picked as  
internal nodes of the binary trees.  This can be achieved using the proposed
\modbtree communication pattern.

\subsection{Impact on performance}

In this section, we assess the impact of using different tree-based 
restricted collective communication schemes on the overall 
performance of \pselinv, and compare the strong scaling of 
\pselinv using either (1) \flattree,
(2) \btree
and (3) \modbtree communication patterns. 


The new implementation of \pselinv contains additional code
improvements that do not fall into the scope of this paper. Therefore,
to emphasize the impact of the different implementations of 
restricted collective communication only, we use the wallclock timing 
measurements of the new \pselinv using the \flattree approach
as the baseline for comparison.
We also provide timings from our previous v0.7.3 release of
\pselinv~\cite{JacquelinLinYang2014} for reference.
This implementation also uses a \flattree communication pattern. 
The wallclock time of the $LU$ factorization based on
\superlu~\cite{LiDemmel2003} is also provided. This is a pre-processing
step of \pselinv. The \superlu timing results are also used here as a
reference for evaluating the strong scaling \pselinv.

Every data point generated from the strong scaling experiments presented 
in this section corresponds to the average of 6 runs.
We report standard deviations by using error bars. We define 
the speedup factor and other ratios as the ratio between average values.

\begin{figure*}[htbp]
\centering
\subfloat[DG\_PNF14000]{
\label{fig:numerDG}
\begin{adjustbox}{width=.45\linewidth}
  \includegraphics{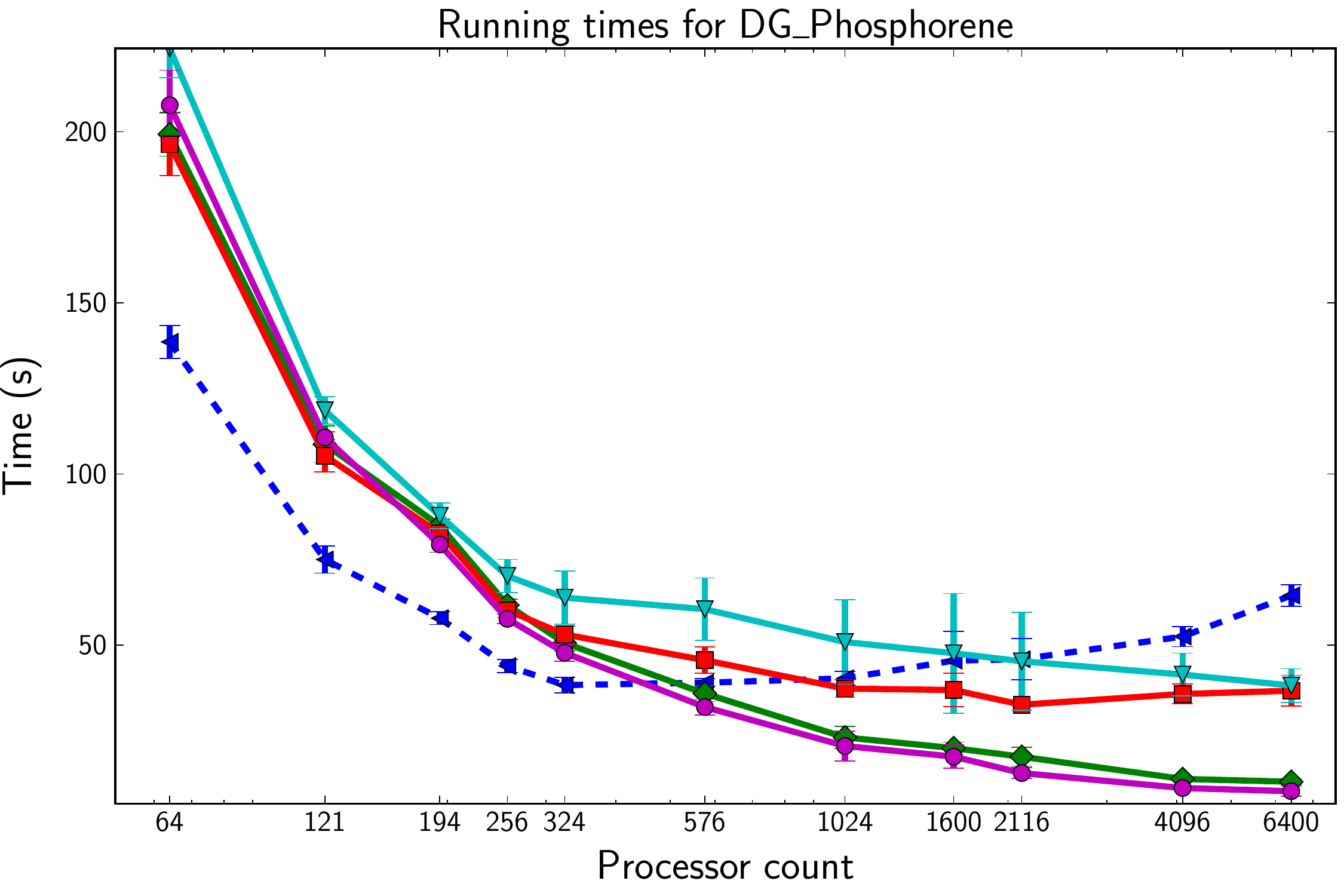}
\end{adjustbox}
}
\subfloat[audikw\_1]{
\label{fig:numerAudi}
\begin{adjustbox}{width=.45\linewidth}
  \includegraphics{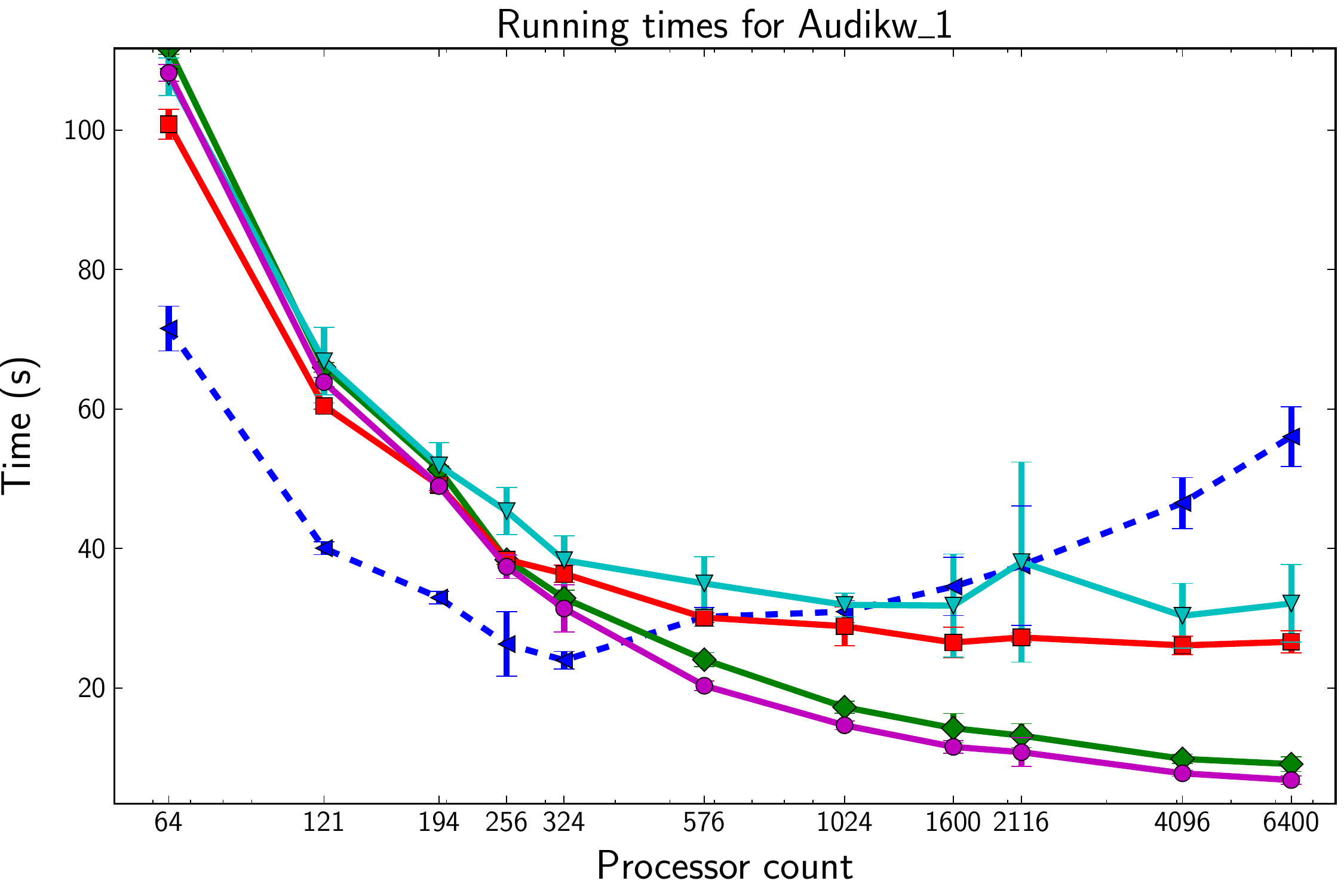}
\end{adjustbox}
}

\subfloat{
\begin{adjustbox}{width=.80\linewidth}
  \includegraphics{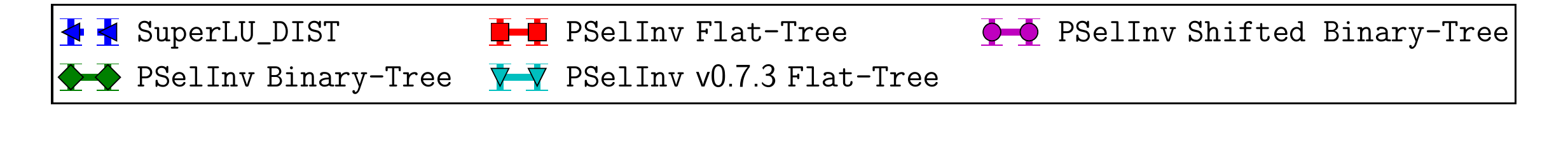}
\end{adjustbox}
}

\caption{Running times of \pselinv for two sample matrices }
\end{figure*}


Results for the DG\_PNF14000 matrix are depicted in Figure~\subref*{fig:numerDG}.
We observe that switching from the \flattree to the \btree scheme 
leads to a reduction of the wall clock time by a factor of 1.7 on average.
The reduction factor is larger when a larger number of processors 
are used in \pselinv. In particular, when more than 1,024 processors
are used, the average speedup factor is 2.5.  The speedup factor 
reaches 3.6 when the computation is performed on 6,400 processors.
Additional performance improvement can be seen when \modbtree 
scheme is used. In particular, the average speedup factor is 
increased to 2.0. When more than 1,024 processors are used, the
average speedup factor increases to 3.2. The maximum speedup reaches 
5.0x when 6,400 processors are used.

Switching from \flattree to \btree also reduces the standard deviation 
of the wall clock time among multiple runs of the same code on the 
same input by a factor of 1.35 on average. The reduction factor is 
1.26 when the \modbtree is used. Compared to v0.7.3 of \pselinv presented
in~\cite{JacquelinLinYang2014}, the average reduction in standard deviation 
resulting from the use of \modbtree is more than 3.19.  

Similar observations holds for the audikw\_1 matrix.  The strong scaling plot
depicted in Figure~\subref*{fig:numerAudi} demonstrates the use of \btree and 
\modbtree allows us to scale the computation to 6,400 processors, whereas
the scalability of \flattree based \pselinv calculations is limited to less 
than 1,024 processors.
The standard deviation in running time is reduced by more than a factor of 4
when a large number of processors are used to run the same program with 
the same input multiple times.

\begin{figure}[h]
\centering
\subfloat[\flattree]{
\label{fig:comp_comm_costs_ftree}
\begin{adjustbox}{width=.43\linewidth}
  \includegraphics{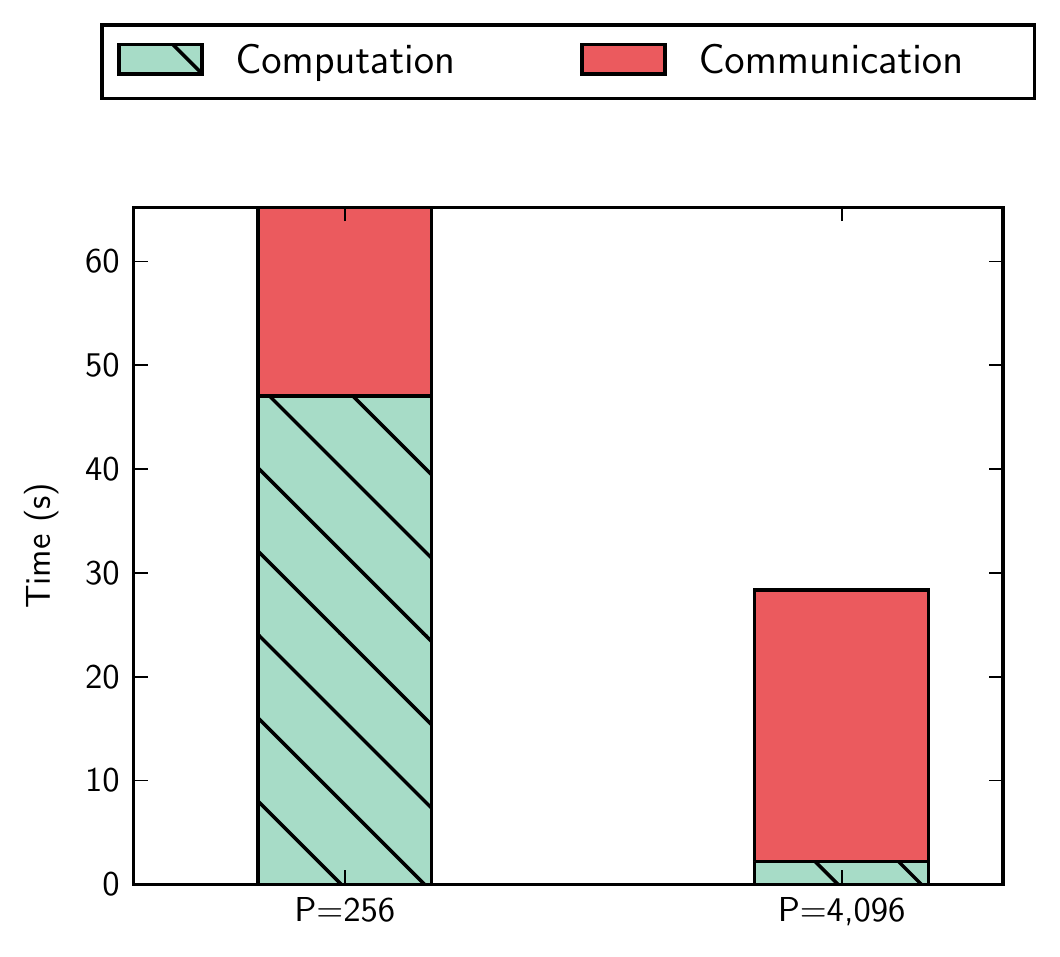}
\end{adjustbox}
}
\subfloat[\modbtree]{
\label{fig:comp_comm_costs_modbtree}
\begin{adjustbox}{width=.43\linewidth}
  \includegraphics{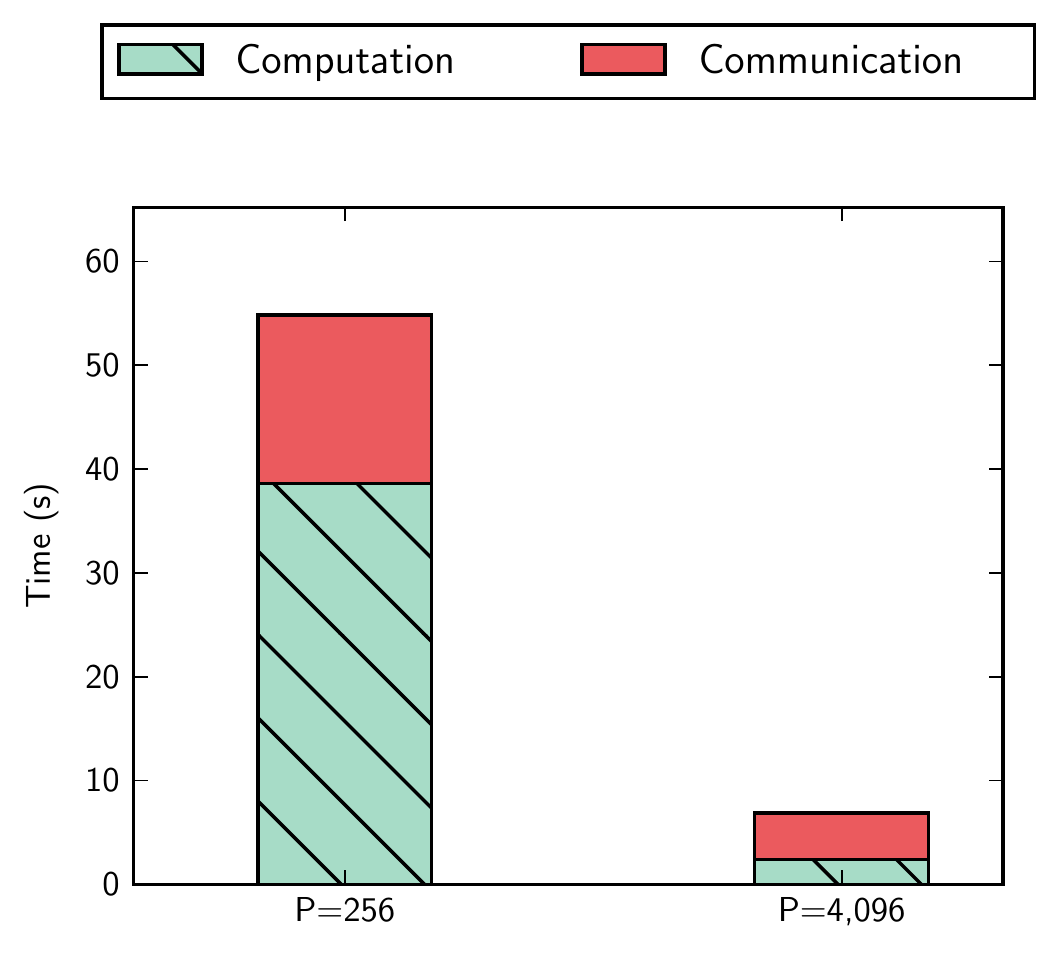}
\end{adjustbox}
}
\caption{Computation and communication times for DG\_PNF14000 }
\label{fig:comp_comm_costs}
\end{figure}

The improved scalability of \pselinv clearly results from a better 
implementation of restricted collective communication which 
significant reduces communication overhead. This can also 
be seen from the ratio of computation and communication time. 
In Figures~\subref*{fig:comp_comm_costs_ftree} and~\subref*{fig:comp_comm_costs_modbtree}, we plot both the communication and computation time 
consumed by \pselinv for the DG\_PNF14000 matrix when the computation
is carried out on 256 and 4,096 processors respectively.
The communication to computation ratio is reduced from 11.8 to 1.9 
when we switch from a \flattree based communication scheme a 
\modbtree based scheme.


It is interesting to note that in both test cases, 
the benefit of using \modbtree is not so pronounced when
the \pselinv is carried out among a small set of processors
(e.g. 256). This is due to the fact that in this case,
several restricted collective communications take place 
within a single node of Edison, which has 24 cores.
Because message passing is implemented as memory copies within 
shared memory on a single node, its cost is generally lower
compared to internode communication. Moreover, having a single 
send buffer in a \flattree based scheme could enhance cache reuse
and reduces the impact of issuing $p$ messages
compared to the $\log_2 p$ messages sent by a binary tree based
collective communication scheme.
Therefore, in practice, one can potentially use a ``hybrid'' 
scheme in which a \flattree based collective communication
is used when the communication is restricted to a relatively
small number of processors and a \modbtree based 
scheme is used when a large number of processors are involved.

\section{Conclusion and future work}\label{sec:conclusion}

We described several implementations of restricted collective 
communication in a parallel selected inversion algorithm.
Each implementation uses the point-to-point \texttt{MPI\_Isend} 
and \texttt{MPI\_Irecv} functions available in a standard MPI 
library. However, they differ in the way each message is 
moved from one processor to another. We showed that
a binary tree based data propagation scheme is far superior
than a flat tree based scheme when a large number of processors
are involved in the collective communication. In particular,
the binary-tree based scheme minimizes communication load 
imbalance and removes communication ``hot spots''. 

The use of \texttt{MPI\_Isend} and \texttt{MPI\_Irecv} 
allows multiple collective communications to be initiated
at the same time. This is a desired feature that would allow
us to exploit a higher level of concurrency in the \pselinv 
algorithm. In order to prevent a processor from becoming 
an internal node of multiple binary trees, we developed
a heuristic that involves applying a random circular shift to 
the list of receiving processor ranks. We demonstrated that such
a heuristic leads to a significant improvement in the scalability
of \pselinv. For instance, when $6,400$ processors are used, we observe
over 5x speedup for test matrices. It also reduces the variation in running time when
the same program is executed multiple times with the same input.
Such variation is caused by inhomogeneous network architecture.
Reducing this type of variation is extremely important for 
achieving scalable performance of the PEXSI
algorithm~\cite{LinLuYingCarE2009,LinYangLuEtAl2011,LinChenYangEtAl2013}
in which multiple selected inversions are carried out simultaneously on
different subgroups of processors. Although our implementation in this
work is for symmetric matrices, the same communication strategy can
be naturally extended to asymmetric matrices, which is our work in
progress.

\section*{Acknowledgment}
This work was partially supported by the Scientific Discovery through
Advanced Computing (SciDAC) program (M. J., L. L.  and C. Y.), and the
Center for Applied Mathematics for Energy Research Applications (CAMERA)
(L. L. and C. Y.), which are partnerships between Basic
Energy Sciences (BES) and Advanced Scientific Computing Research (ASCR)
at the U.S Department of Energy.

\bibliographystyle{unsrt}
\bibliography{pselinv}

\end{document}